\documentclass{emulateapj}
\usepackage{apjfonts}
\slugcomment{Accepted to ApJ, August 6 2009; published October 1 2009} 
\usepackage{natbib}
\usepackage{natbib,natbibspacing}
\shorttitle{LBGs in the UDF}
\shortauthors{Rafelski, Wolfe, Cooke, Chen, Armandroff, \& Wirth}
 
\begin{document}

\title{Deep Keck $u$-band imaging of the Hubble Ultra Deep Field: \\ A catalog of $z\sim3$ Lyman Break Galaxies \\}

\author{Marc Rafelski,\altaffilmark{1} Arthur M. Wolfe,\altaffilmark{1} Jeff Cooke,\altaffilmark{2} Hsiao-Wen Chen,\altaffilmark{3} Taft E. Armandroff,\altaffilmark{4} \& Gregory D. Wirth\altaffilmark{4}}
\email{marcar@ucsd.edu} 
\altaffiltext{1}{Department of Physics and Center for Astrophysics and Space Sciences, UCSD, La Jolla, CA 92093, USA}
\altaffiltext{2}{The Center for Cosmology and the Department of Physics and Astronomy, UCI, Irvine,CA 92697, USA}
\altaffiltext{3}{Department of Astronomy and Astrophysics, University of Chicago, Chicago, IL 60637, USA}
\altaffiltext{4}{W. M. Keck Observatory, Kamuela, HI 96743, USA}

\begin{abstract}

We present a sample of 407 $z\sim3$
Lyman break galaxies (LBGs) to a limiting isophotal $u$-band magnitude of 27.6 mag
in the Hubble Ultra Deep Field (UDF).  The LBGs are
selected using a combination of photometric redshifts and 
the $u$-band drop-out technique enabled by the
introduction of an extremely deep $u$-band image obtained with the
Keck I telescope and the blue channel of the Low Resolution Imaging Spectrometer.
The Keck $u$-band image, totaling 9 hrs of integration time,
has a 1$\sigma$ depth of $30.7$ mag arcsec$^{-2}$, making it one of
the most sensitive $u$-band images ever obtained.  The $u$-band
image also substantially improves the accuracy of photometric redshift
measurements of $\sim50\%$ of the $z\sim3$ LBGs, 
significantly reducing the traditional degeneracy of colors between $z\sim3$ and 
$z\sim0.2$ galaxies. 
This sample provides the most
sensitive, high-resolution multi-filter imaging of reliably identified
$z\sim3$ LBGs for morphological studies of galaxy formation and
evolution and the star formation efficiency of gas at high redshift.

\end{abstract}


\pagebreak
\clearpage

\section{Introduction}

The Hubble Ultra Deep Field \citep[UDF;][]{Beckwith:2006p1529} provides the 
most sensitive high-resolution images ever taken, 
yielding a unique data set for studying galaxy evolution. 
These data have contributed to many scientific advances, including 
constraining the star formation efficiency 
of gas at $z\sim3$ \citep{Wolfe:2006p474}, aiding in determining the
luminosity function in the redshift range $4\lesssim z \lesssim 6$
 \citep{Bouwens:2007p4335}, bringing insight into 
the merger fractions of galaxies \citep{Conselice:2008p5047}, 
and yielding the discovery of clumpy
galaxies at high redshift \citep{Elmegreen:2007p1537}. 

Knowledge of galaxy redshifts is essential to understanding their
nature, and various approaches are used to estimate this key
attribute.  A number of studies identify objects in the redshift range
$4\lesssim z \lesssim 6$ using the deep multi-filter (BVIZJH) UDF images 
by identifying so-called ``dropout" galaxies 
which are detectable in certain broadband filters but not in others 
\citep{Beckwith:2006p1529, Bouwens:2006p4586, Bouwens:2007p4335}.
Others use photometric redshifts derived across the entire redshift range
$0 < z \lesssim 6$ by analyzing the colors of galaxies in a wider range of 
filters \citep{Coe:2006p1519}.  This latter approach has the potential to provide 
relatively accurate \citep[$\sigma_{\Delta z/(1+z)} \lesssim 0.1$, ][]{FernandezSoto:2001p2773} 
redshift estimates for a large number of galaxies in a
given field, but traditionally has serious problems producing accurate
results near $z\sim3$, a redshift range for which
key spectral energy distribution (SED) features fall
blueward of the previously available filter set.

The majority of galaxies observed at $z\sim3$ are Lyman break galaxies (LBGs): star-forming galaxies that are
selected based on the break in their SED at the 912~\AA~Lyman limit primarily by 
interstellar gas intrinsic to the galaxy, as well as a flux decrement shortward of 1216~\AA~in the galaxy rest frame due to 
absorption by the Lyman series of opticallythick hydrogen gas along the line of sight.
This Lyman limit discontinuity in the SED 
significantly dims these galaxies shortward of $\sim3500$~\AA, allowing them to be found
with the U-band dropout technique \citep{Steidel:1992p1911,Steidel:1995p1873, Steidel:1996p5981, Steidel:1996p5985}. 
Previously, the available observations in the 
UDF did not include deep $u$-band imaging. The 
next prominent broadband signature in the SED is the 
4000~\AA~break, which is redshifted to the near infrared (IR) for $z\sim3$ LBGs.
Without the $u$-band or very deep IR coverage, it is very difficult to determine from broadband imaging alone
whether the observed decrement in the SED near the observed-frame $\sim4800$~\AA~is
a low-redshift galaxy with a 4000~\AA~break, or a high-redshift galaxy with a decrement from 
the 1216~\AA~break. This degeneracy causes ``catastrophic" errors 
in the photometric redshifts of galaxies at $z\sim3$ without $u$-band data 
\citep{Ellis:1997p3771, FernandezSoto:1999p2784, Benitez:2000p3572}. 

The purpose of this paper is to present a reliable sample of 
LBGs at $z\sim3$ in the UDF through the introduction of 
ultra-deep u'-band imaging acquired  with the 10m Keck I telescope. 
The Keck I telescope and the Low Resolution Imaging Spectrometer \citep[LRIS;][]{Oke:1995p6046,McCarthy:1998p6102}
form an ideal combination to probe galaxies at the $z\sim3$ epoch due to the
light-gathering power of the 10 m primary mirror and
the outstanding efficiency of the LRIS blue channel in the near UV.
This allows the $u$-band filter (with effective wavelength $\lambda_{o} \sim3400$~\AA, FWHM$\sim690$~\AA) used with the blue arm (LRIS-B) 
to be $\sim300$~\AA~bluer and $\sim360$~\AA~wider than the $u$-band filter on the 
Visible Multi-Object Spectrograph \citep[VIMOS;][]{LeFevre:2003p8774}
instrument ($\lambda_{o} \sim3700$~\AA, FWHM$\sim330$~\AA) at the Very Large Telescope (VLT) and still be effective. 
Consequently, this enables us to probe the Lyman break efficiently to a lower limit of $z\sim2.5$, versus the VIMOS limit of $z\sim2.9$
\citep[as shown by the VIMOS observations of GOODS-South by][]{Nonino:2009p9926}.
In addition, although the UDF field can only be observed at high air mass from Mauna Kea,
the total throughput of LRIS-B and its bluer $u$-band is approximately twice that of VIMOS with its redder $u$-band.
This gives Keck the unique ability to select lower redshift LBGs via their Lyman break. 

We assemble a reliable sample of LBGs in the UDF with a combination of the $u$-band 
dropout technique and photometric redshifts, and provide photometric redshifts for all 
galaxies with good $u$-band photometry. We find that
the $u$-band imaging improves the photometric redshifts
of $z\sim3$ galaxies by reducing the degeneracy between low and
high-redshift galaxies. Furthermore, the combination of deep LRIS $u$-band and
high-resolution multiband Hubble Space Telescope (HST) imaging provides an unprecedented view of $z\sim3$ LBGs
to improve our understanding of their highly irregular rest-frame UV morphologies \citep{Law:2007p5043}
down to unprecedented depths. 

The deep, high-resolution LBG sample will also extend current constraints
on the star formation efficiency of gas at $z\sim3$. 
Reservoirs of neutral gas are needed to provide the fuel for star formation.  
Damped Ly${\alpha}$ systems (DLAs), selected for their neutral hydrogen column 
densities of $N_{\rm H I} \geq2\times10^{20}$cm$^{-2}$, dominate the neutral-gas content of the universe in the
redshift interval $0<z<5$. DLAs contain enough gas to account for 50\% of the 
mass content of visible matter in modern galaxies \citep[see][for a review]{Wolfe:2005p382} and may
act as neutral-gas reservoirs for star formation, since stars form when local values
of the ${\rm H~I}$ column density, $N_{\rm H I}$, exceed a critical value. 
At high redshift, the star formation rate (SFR) is assumed to follow
the Kennicutt-Schmidt (KS) law (established for nearby galaxies) which states how the SFR 
per unit physical area, $\dot{\Sigma_{*}}$, relates to the neutral gas (i.e., ${\rm H~I}$ and ${\rm H_2}$) column density:
$\dot{\Sigma_{*}} \propto N_H^{1.4}$ \citep{Kennicutt:1998p3174}.
Strong DLAs with $N_{\rm H I} \geq1.6\times10^{21}$cm$^{-2}$ in the redshift range
$2.5\lesssim z \lesssim 3.5$ are predicted to have emission from star formation in the rest-frame
FUV redshifted into the optical such that they are detectable in the UDF F606W image.
\citet*{Wolfe:2006p474} searched for low surface-brightness
emission from DLAs in the UDF and found the {\it in situ} SFR efficiency of DLAs to 
be less than 5$\%$ of the KS law. In other words, star formation must occur 
at much lower rates in DLAs at $z\sim3$ than in modern galaxies. 
Whereas the \citet{Wolfe:2006p474}
results set sensitive upper limits on  {\it in situ}  star formation
in DLAs excluding known galaxy regions, no such limits exist for DLAs containing LBGs.  
The sample presented here enables a search for 
spatially extended low-surface-brightness emission around $z\sim3$ LBGs 
which will yield constraints on the star formation efficiency at high redshift (Rafelski et al. in prep). 
This is one of the main motivations for constructing this sample, and therefore we 
construct our sample conservatively to minimize the number of potential interlopers.

We present our catalog of $z\sim3$ LBGs, provide photometric redshifts for the entire sample of objects 
that have reliable  $u$-band photometry, and make the $u$-band image available to the public. 
The observations are described in \S2, and
the data reduction and analysis in \S3.  We discuss the
photometric selection of  $z\sim3$ galaxies in \S4, and summarize our major findings in \S5.
Throughout this paper, we adopt the AB magnitude system and an 
$(\Omega_M, \Omega_\Lambda, h)=(0.3,0.7,0.7)$ cosmology with parameters $\Gamma=0.21, n=1$, 
which are largely consistent with recent values \citep{Hinshaw:2009p8215}.

\begin{figure*}
\begin{center}
\plottwo{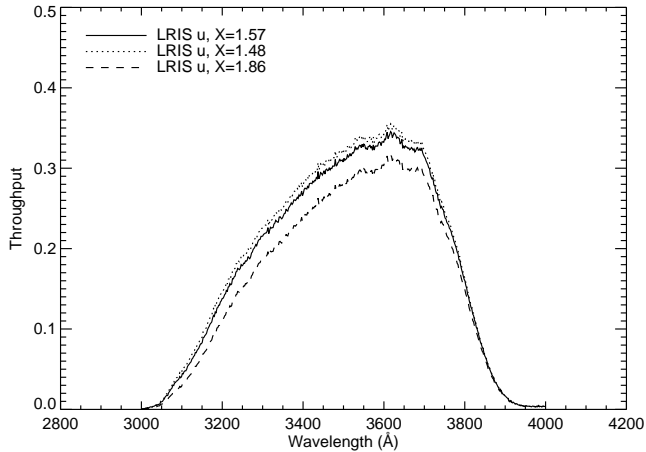}{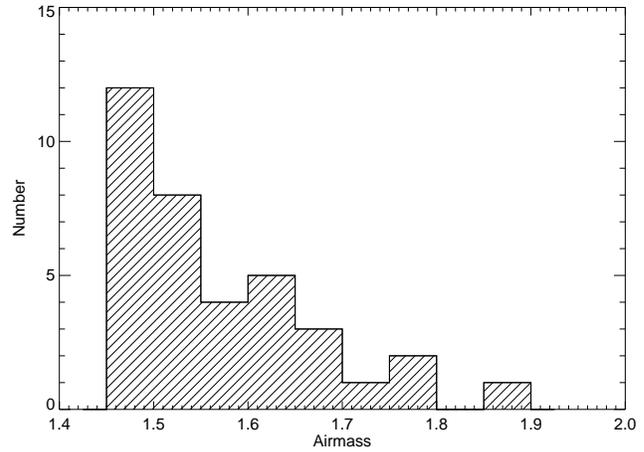}
\caption{Transmission of the $u$-band filter used in this study for different air masses ($X$). The filters 
are corrected for the atmosphere attenuation at different air masses and the CCD quantum efficiency of LRIS-B on the Keck telescope. 
The histogram shows the range of air masses in the study, with the minimum and maximum air mass being plotted on the transmission curve. 
The change in the effective wavelength $\lambda_{o}$ between air masses is typically $\lesssim 5$~\AA.
\label{airmass}}
\end{center}
\end{figure*}

\section{Observations}

The $u$-band ($\lambda_{o} \sim3400$~\AA, FWHM$\sim690$~\AA) images of the 
UDF ($\alpha(J2000) = 03^{h}32^{m} 39^{s}$, $\delta(J2000) = -27^{\circ}$47$\tt'$29.$\tt''1$)
were obtained with the 10 m Keck I telescope using the LRIS.
The $u$-band data were taken with the new Cassegrain Atmospheric Dispersion Corrector \citep[Cass ADC;][]{Phillips:2006p6047}
to minimize image distortions from differential atmospheric refraction.
The Cass ADC was critical to the success of these observations because of the low elevation of the UDF from Mauna Kea,
and the blue wavelength of our primary band. 
The data also benefit from the backside illuminated, dual
UV-optimized Marconi $2048\times4096$ pixel CCDs on LRIS-B,
with 0.$\tt''$135 pixels and very high UV quantum efficiency ($\sim$ 50\% at $\lambda_{o} \sim3450$~\AA).  
Because the quantum efficiency of the two CCD chips varies $\sim$30\%--35\% in the $u$-band,
we placed the UDF entirely on the more sensitive chip (CCD1).
We used a dichroic beam splitter (D460) to simultaneously observe the $u$-band on the blue side, 
and the $V$-band and $R$-band on the red side. 
The red channel data were taken for astrometric and photometric consistency checks, 
and were not used in this study other than for calibration purposes
due to the much deeper and higher resolution UDF images available over those wavelength ranges.

The observations were carried out in darktime over two runs,
2007 October 7--9 (three half nights) and 2007 December 3--4 (two half nights). 
We lost the entire first night of the first run to weather, and had moderate weather and seeing conditions 
throughout both runs that yielded a median seeing FWHM of $\sim$1.$\tt''$3 in the $u$-band.
In order to maximize time on sky, we adopted the dither strategy of \citet[see their Eqn. 1]{Sawicki:2005p1714}, setting the red
channel exposure times such that the last readout of the red channel coincided with the end 
of the blue channel readout. 
The $u$-band images were acquired as a series of $36\times900$s exposures to avoid the nonlinear regime of the CCD, 
totaling 9 hrs of integration time on target.
We executed a nine-point dither pattern with 10$\tt''$ dithers
to deal with bad pixels and to create a super-sky flat. 

The UDF can only be observed at large air masses at Mauna Kea, which can affect the shape of the $u$-band throughput.
We show a histogram of our observed air masses and the variation of the filter throughput for different air masses  in Figure \ref{airmass}.
We find that the variations in air masses in our sample do not significantly affect the blue side cutoff of our 
filter.  Specifically, we find that the variation between our best and worst air mass yields a change in 
$\lambda_{o} \lesssim 10\AA$, with typical changes in $\lambda_{o}$ being $\lesssim 5\AA$.
Given the small variation of $\lambda_{o}$, we derive the final $u$-band filter transmission curve by convolving the 
measured filter throughput with the atmospheric attenuation at our average air mass of 1.57 and the CCD quantum efficiency.

\begin{figure}
\begin{center}
\plotone{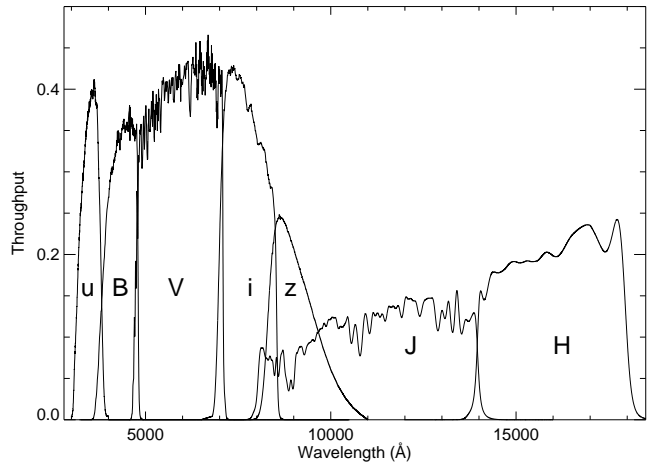}
\caption{Transmissions of the filters used in this paper, corrected for the CCD quantum efficiency, 
and in the case of the $u$-band, the atmosphere attenuation at the average air mass of 1.57. 
The $u$-band is from LRIS-B on the Keck telescope, 
the $B$, $V$,  $i^\prime$, and $z^\prime$ bands from the WFC on the HST ACS, and the $J$ and $H$ bands from the NICMOS camera on HST.
\label{filter}}
\end{center}
\end{figure}

Throughout the paper we utilize the $B$, $V$, $i^\prime$, and $z^\prime$ band 
(F435W, F606W, F775W, and F850LP, respectively) observations of the 
UDF  \citep{Beckwith:2006p1529}, obtained 
with the Wide Field Camera (WFC) on the HST Advanced Camera for Surveys \citep[ACS;][]{Ford:2002p6197}.
These images cover 12.80 arcmin$^{2}$, although we prune our catalog to 
the central 11.56 arcmin$^{2}$ which contains at least half the average depth of the whole image
and overlaps our $u$-band image with uniform depth. In addition to these ACS images, 
we also include observations taken with the NICMOS camera NIC3 in the $J$ and $H$ bands 
\citep[F110W and F160W;][]{Thompson:2006p1569}. These red wavelengths cover
the central 5.76 arcmin$^{2}$ of the UDF, so we only use them whenever the field of view (FOV) overlaps.
Figure \ref{filter} plots the total throughput of the filters used in this paper:
the Keck LRIS-B $u$-band, the HST ACS $B$, $V$,  $i^\prime$, and $z^\prime$ bands, 
and the HST NICMOS $J$ and $H$ bands, and include the CCD quantum efficiency and atmospheric attenuation.

\vspace{0.2in}

\section{Data Reduction and Analysis}

\subsection{Image processing}
The LRIS-B data were processed in a combination of custom code in IDL,
and standard data reduction algorithms from IRAF\footnote{
IRAF is distributed by the National Optical Astronomy Observatory, 
which is operated by the Association of Universities for Research in Astronomy,
Inc., under cooperative agreement with the National Science Foundation.}.
The images were bias subtracted, first from the overscan region and then
from separate bias frames to remove any residuals, before being trimmed 
to remove any vignetted regions of LRIS.
Super-sky flats  were created using IRAF from the median of all the unregistered images with 
sigma clipping to remove objects and cosmic rays, which were then used to flat-field the images.
These prove superior to dome and twilight sky flats in determining the CCD response in the $u$-band
because of the short wavelengths, and yield excellent flats.
Dithering offsets were determined using custom code and $\tt SExtractor$ \citep{Bertin:1996p6133} 
to locate bright objects common to all the images. 
In order to {\it drizzle} \citep{Fruchter:2002p6141} only once and keep correlated noise to a minimum, 
the images were distortion corrected using the solution provided 
by J. Cohen \& W. Huang (private communication, February 2008) and shifted to 
correct for the dithering offsets all at once using the $\tt geotran$ package in IRAF.
We drizzled with $\tt pixfrac = 0.5$ to improve the point-spread function (PSF) and set
the pixel scale such that the pixels are integer multiples of the UDF pixels, 0.$\tt''$12, for reasons explained in \S3.4.
To maximize the signal to noise (S/N), the images were weighted by their inverse variances. 
The drizzled images were then combined using the IRAF task $\tt imcombine$, with sigma clipping to remove bad pixels and cosmic rays.
These images were trimmed to include regions of uniform depth (11.56 arcmin$^{2}$), normalized to an effective exposure time of 1s, and then background subtracted 
using the global background determined with $\tt SExtractor$.
An astrometric solution was applied with the IRAF task $\tt ccmap$ by matching bright and nearly 
unresolved objects in the UDF to those in the final stacked image. 
The final rms astrometric errors are between 0.$\tt''$02 and 0.$\tt''$03, 
negligible in comparison to the 1.$\tt''$3 FWHM of the $u$-band PSF.

\subsection{Photometric Calibration}

Moderate weather yielded no completely photometric night for our calibration, and 
we therefore calibrated to the Multi-wavelength Survey by Yale--Chile \citep[MUSYC;][]{Gawiser:2006p2052}, which covers the 
same part of the sky as our primary observations. We include all our filters ($u$, $V$, and $R$) for this calibration. 
All our calibrations are in the AB95 system of \citet{Fukugita:1996p2320}, hereafter referred to as AB magnitudes.
We used the IRAF tasks $\tt phot$ and $\tt fitparams$ to solve for zero-point magnitudes, 
air-mass correction coefficients, and appropriate color correction coefficients. Specifically, we use the equation:
\begin{equation}
\rm m = -2.5 \log_{10}(F) + Z - c X - Y,
\label{eq:photcal}
\end{equation}
where $F$ is the flux in counts/s, $Z$ is the zero-point magnitude, $c$ is the air-mass coefficient, $X$ is the air-mass, and $Y$ is the color term. We allow a color term 
to account for any differences between the $U$-band filter used by MUSYC and the $u$-band filter in our observations, 
similar to the color term used by \citet{Gawiser:2006p2052} to correct for their differences compared to the Johnson-Cousins filter set. 
The galactic extinction of 0.0384 is subtracted from the zero-point magnitude, using the relation
$A(u)=4.8E(B-V)$ interpreted from \citet{Cardelli:1989p2011}, where $E(B-V)=0.008$ \citep{Beckwith:2006p1529}.
The final results for the $u$-band calibration are a zero-point magnitude $Z=27.80\pm0.03$, an air-mass coefficient term $c=0.41$, an average air-mass  $X=1.57$, 
and a color term $Y=(0.13\pm0.02)\times(U-B)_{AB}$.
We double check our calibration using  multiple observed photometric standard stars \citep{Landolt:1992p6144} 
over a range of air masses, and the zero-point and air-mass correction coefficients are consistent
with those found when calibrating to the MUSYC catalog. 
As a result, we are confident that the $u$-band image is well calibrated.

\subsection{Depth of the $u$-band Image}

It is useful to characterize the depth of the $u$-band image, however,
different definitions exist to describe the sensitivity of an image. Two commonly quoted limits
are presented here: a measurement of the
sky fluctuations of the image, and a limiting magnitude corresponding to a 50\% decrease in object counts
through Monte Carlo simulations. 

The sky noise of the image is measured via the pixel to pixel rms fluctuations in the image, best measured by fitting a Gaussian to
the histogram of all pixels without sources. Sources are identified with the program $\tt SExtractor$, with the threshold set
such that the negative image has no detections. This yields a depth of 31.0 mag arcsec$^{-2}$, $1\sigma_{u}$ sky fluctuations.
However, since the image is drizzled, correlated noise between the pixels 
is introduced. The theoretical increase in noise due to $\tt pixfrac = 0.5$ using equation 10 from 
\citet{Fruchter:2002p6141} is 20\%. Alternatively, the correlated noise can be estimated 
empirically with equation 2 from \citet{FernandezSoto:1999p2784}, which uses a covariance matrix to determine
a small overestimate of the real error\footnote{The equation has a typographic error. 
The sum should be over $i_1,i_2=1$ to $i_1,i_2=3$.}.
This results in a slightly more conservative depth of 30.7 mag arcsec$^{-2}$, $1\sigma_{u}$ sky fluctuations,
which is what we quote here.

\begin{figure}
\begin{center}
\plotone{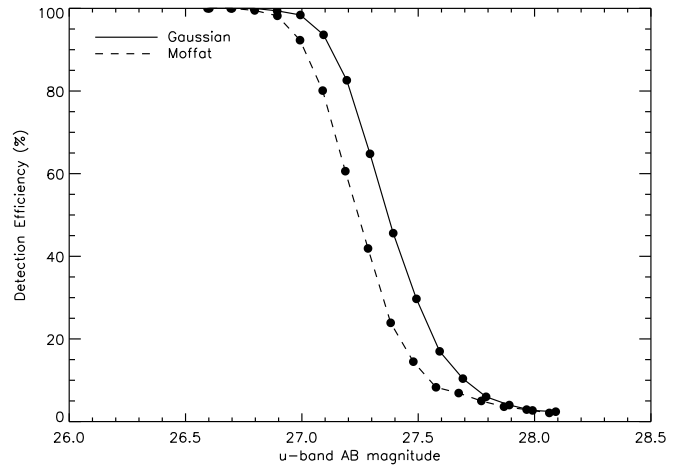}
\caption{Detection efficiency of the u image based on simulations planting objects of different intrinsic profiles into the image. The solid line represents a gaussian profile with a FWHM of  1.$\tt''$3 and the dashed line represents a Moffat profile with a half-width at half maximum of 1.$\tt''$2 and a power law index of 3.4.
The u values are total magnitudes based on $\tt SExtractor$'s $\tt mag\_auto$ apertures which are Kron-like \citep{Kron:1980p9237} elliptical apertures corrected for possible contamination.
We find a total $u_{\mathrm{lim}}$ of 27.3 mag for the gaussian and 27.2 mag for the Moffat profile, and for isophotal apertures which are more appropriate for LBGs, find 
a $u_{\mathrm{lim}}$ of 27.7 mag for the Gaussian and 27.6 mag for the Moffat profile.
\label{det}}
\end{center}
\end{figure}

Additionally, to get a better sense of the usable depth of the image, 
a limiting magnitude is often quoted \citep[e.g.,][]{Chen:2002p2597, Sawicki:2005p1714}.
We define $u_{\mathrm{lim}}$ as the magnitude limit at which more than $50\%$ of the objects are detected.
The best way to determine $u_{\mathrm{lim}}$
is through Monte Carlo simulations, which take into account both the sky surface brightness and the seeing in our image. 
Since our image PSF has more flux in the wings than a Gaussian, we plant both Gaussian objects, and objects modeled to fit
our PSF using a two-dimensional (2-D) Moffat profile\footnote{The Moffat profile \citep{Moffat:1969p7422} is a modified Lorentzian with a 
variable power-law index that takes into account the flux in the 
wings of the intensity profile which are not included in a Gaussian profile.}. 
The custom IDL code was used to
extract bright unresolved objects in the $u$-band image, take the median of all 
these objects, and create a composite object stack. The composite image was then 
fit both by a Gaussian, and by a 2-D Moffat profile using $\tt MPFIT$ \citep{Markwardt:2009p7396}, 
with a modification to ensure that the wings of the profile go to 0 for the Moffat profile.
We semi-randomly insert these objects with a range of fluxes into the $u$-band image. 
The locations of the planted objects are constrained such that they do not: 1) fall off the edges,
2) fall on a real detected object, and 3) fall on any previously planted objects. 
We find a total $u_{\mathrm{lim}}$ of 27.3 mag for the Gaussian and 27.2 mag for the Moffat profile (see Figure \ref{det}), 
where the total magnitudes are based on $\tt SExtractor$'s $\tt mag\_auto$ apertures, which 
are Kron-like \citep{Kron:1980p9237} elliptical apertures corrected for possible contamination. 
While total magnitudes are generally reported, 
isophotal apertures are more appropriate for LBGs, and yield a $u_{\mathrm{lim}}$ of 27.7 mag for the 
Gaussian and 27.6 mag for the Moffat profile.
This is one of the deepest $u$-band images ever obtained, with our sensitivity being similar to 
those reported in the Keck Deep Fields \citep{Sawicki:2005p1714}.

This detection method does not match the detection method used in \S3.4, and therefore does not constrain our detection efficiency of LBGs. 
As explained below, we use our prior knowledge of the positions of the sources which yields a different completeness limit. 
However, this result gives the depth of our $u$-band image for comparison to other studies.

\subsection{Photometry through Template Fitting}

In order to obtain robust colors across images with varied PSFs, it is necessary to
match apertures and correct for PSF differences. 
If the difference is minor, then methods to apply aperture corrections to account for 
the variations are appropriate, such as
the $\tt ColorPro$ software by \citet{Coe:2006p1519}. 
However, such algorithms don't perform well 
when the difference in the PSF FWHM is large, such as when combining the high-resolution data from the HST
(0.$\tt''$09 FWHM) with the low-resolution images obtained in this study with Keck (1.$\tt''$3 FWHM). 
In this case, the uncertainties in aperture corrections are unreasonably large, and the low-resolution images are crowded such that 
objects overlap, making object definitions that are valid in both high and low-resolution images difficult to determine.

In order to avoid these uncertainties, we use the $\tt TFIT$  \citep{Laidler:2007p2733} template-fitting method 
that uses prior knowledge of the existence, locations, and morphologies of sources in the deeper high-resolution 
UDF images to improve the photometric measurements in our low-resolution 
$u$-band image. 
This method creates a template of every object by convolving each 
object in the high-resolution image with the PSF of the low-resolution image. 
These templates are then fit to the low-resolution image to determine the flux of all the 
objects in the $u$-band, relative to the flux in the UDF $V$-band image. 
We chose the $V$-band as
the high-resolution reference image because it is closest in wavelength to the $u$-band, 
without being affected by the Ly$-\alpha$ forest over the redshift interval $2.5\lesssim z \lesssim 3.5$. 
The result is a very robust color which relates every object in the high-resolution $V$-band image to
the low-resolution $u$-band, avoiding the problem of aperture matching between the two images while
intrinsically correcting for the PSF difference.
Using the $V$-band flux, the color is converted to a $u$-band flux,  which inherits the same isophotal aperture as the 
high-resolution $V$-band image. The aperture correction used to obtain total fluxes for the $V$-band image
is then also valid to obtain total fluxes for the $u$-band.
For a more in-depth explanation, see \citet{Laidler:2007p2733}.
This technique is similar to others in the literature \citep{FernandezSoto:1999p2784, Labbe:2005p6209, 
Shapley:2005p4909, Grazian:2006p6244},
with the original version based on \citet{Papovich:2001p4910, Papovich:2004p6306}. 

We chose to use $\tt TFIT$ as it is publicly available, well documented, and its performance is carefully tested. 
Like all the other methods, there are some constraints that had to be met to 
use this algorithm. The first is that the pixel scale of the $u$-band image must be 
an integer multiple of the pixel scale of the UDF $V$-band image. 
This was accomplished by drizzling the images such that 
the pixel scale of the $u$-band is 4 times larger than the $V$-band, as mentioned in \S3.1. 
The second and third requirements are that the images must not be rotated with respect to each 
other, and the corner of the $V$-band image must coincide with the corner of the $u$-band 
image. These two requirements were met
by rotating and trimming the $V$-band image using IDL. The resultant image was compared to the 
original image, and the difference in photometry was negligible compared to the intrinsic uncertainties. 
We also improved our fit by source weighting the rms map before providing it to the $\tt TFIT$ pipeline, as suggested by \citet{Laidler:2007p2733}. 

In order to avoid proliferating different catalogs with minor differences in object definitions, 
we adopt the object definitions of \citet{Coe:2006p1519}. These definitions include the catalogs 
of \citet{Beckwith:2006p1529} and \citet{Thompson:2006p1569}, as well as detections performed on a white
light image by \citet{Coe:2006p1519}. By using identical object definitions as \citet{Coe:2006p1519}, we can use the 
careful photometry for the $B$, $V$,  $i^\prime$, $z^\prime$,  $J$, and $H$ bands already determined, and compare
our redshift determinations knowing we have used identical apertures. 

$\tt TFIT$ requires a $\tt SExtractor$ catalog of the $V$-band image as an input to the pipeline.
The program $\tt sexseg$ \citep{Coe:2006p1519} was run on the segmentation map of \citet{Coe:2006p1519} to provide
the necessary information to $\tt TFIT$ while using the desired object definitions of \citet{Coe:2006p1519}. 
The segmentation map defines which pixels belong to each identified $V$-band object. The 
$\tt sexseg$ program forces $\tt SExtractor$ to run using a predefined segmentation map \citep[for details, see][]{Coe:2006p1519}. 
$\tt TFIT$ also requires a representative 2-D model of the $u$-band PSF, and is sensitive to the quality of its construction. 
We use the same Moffat profile fit described in \S3.3 to model the PSF of the $u$-band image, and then use this as
the transfer kernel by $\tt TFIT$ to convolve the $V$-band galaxy cutouts. 
We note that there is no significant spatial variation of the PSF across the field.

In general, the higher the resolution and sensitivity of the high-resolution image, 
the better $\tt TFIT$ can model the sources for the low-resolution image. 
If an input catalog is not complete enough, then unmodeled objects can act 
as an unsubtracted background, slightly increasing the flux of all objects \citep{Laidler:2007p2733}. 
However, this has a limit, and eventually there are so many sources that are too 
faint to detect in the low-resolution image that galaxies 
are not well constrained given the substantial number of priors. This yields 
a large number of galaxies with unconstrained fluxes
that increase the uncertainties of the nearby objects without yielding any new information. 
The UDF $V$-band image has substantially higher resolution and is deeper than the $u$-band image, 
and therefore a limit was put on the faintest galaxy used as a prior in $\tt TFIT$.
Only galaxies brighter than $V=29$ mag are included in the input catalog to $\tt TFIT$.
The galaxies fainter than $V=29$ mag are too faint to be constrained by the u image, and only add noise to the $\tt TFIT$ results. 
We stress that this is a conservative cut, and does not introduce an unsubtracted background. 

\begin{figure*}
\epsscale{0.93}
\plotone{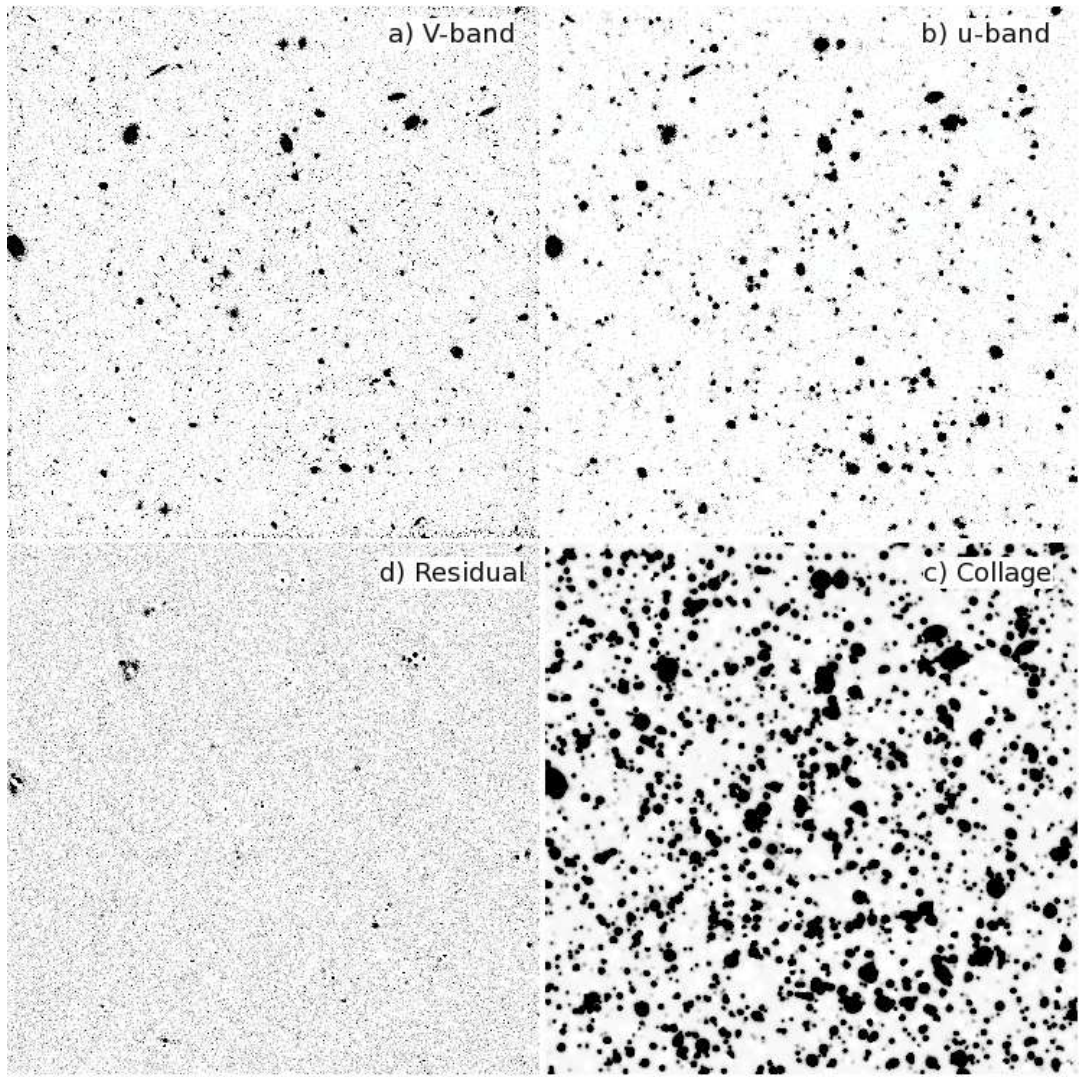}
\caption{Quality of $\tt TFIT$ photometric fits of the UDF $V$-band and
LRIS $u$-band imaging. The images cover the overlap region of the central 
11.56 arcmin$^{2}$ of the UDF, with 
North pointing up and East to the left.  Clockwise from upper left: (a) the $V$-band
image from the UDF; (b) the $u$-band  image from Keck; (c) the model image;
(d) the residual image. The model image is a collage of the $V$-band
galaxies convolved by the PSF of the $u$-band image, scaled by the
$\tt TFIT$ flux measurement for each object. The residual image is the
difference of the model and the $u$-band image. Ideally the residual
image would be zero, but this is not the case for bright objects (see \S3.4).
\label{TFIT}}
\end{figure*}

The quality of the resulting photometric fits can be evaluated through Figure \ref{TFIT}, 
which depicts four panels: the $V$-band image from the UDF, the u image from Keck, 
the model image, and the residual image. The model and residual images are diagnostics produced by 
$\tt TFIT$, and are not used in the fitting process. The model image is a collage of the $V$-band galaxies convolved
with the PSF of the $u$-band image, scaled by the $\tt TFIT$ flux measurement for each object. The residual 
image is the difference of the model and the $u$-band images. Ideally the residual image would be zero, 
but this is not the case (especially for bright objects), with multiple effects contributing to the imperfect residual. 
For instance, if the object in the $V$-band image is saturated, then it has the wrong
profile for the $u$-band and leaves a residual. Alternatively, imperfections in the modeled 
PSFs when scaled to large flux measurements of bright objects will also leave a residual. This effect was 
minimized by using a source-weighted rms map, although photometry of the brightest objects are imperfect.
In practice, it is very difficult to perfectly align two images, the distortion correction is not perfect, and 
images generally have spatially varying PSFs. 
To minimize these affects, $\tt TFIT$ does a ``registration dance", where it cross-correlates each region of the 
model with the region of the data to find any local shifts. 
This registration dance was performed, which slightly improved the residual image, and leads to more robust photometry.

\subsection{Sample Selection}

Our aim is to identify a sample of high-redshift galaxies that are suitable for constraining
the star formation efficiency of gas at $z\sim3$. 
A large fraction of objects for which fluxes are measured with $\tt TFIT$ are too faint to yield sufficient information
regarding the object's redshift, and we therefore limit our sample to those objects with high S/N. 
We select objects based on their $V$-band magnitudes,
since cuts in  $u$-band would preferentially remove LBGs. The median $u$-band S/N of all objects 
decreases as a function of the $V$-band magnitude and drops below 3$\sigma$ at $V\geq27.6$ mag.
We adopt this $V$-band magnitude cut to include the majority of high S/N u objects, 
while removing S/N $<$ 3 objects. We note that this is a conservative 3$\sigma$ cut since most 
LBGs won't be detected in the $u$-band reducing the overall median S/N.

In addition to removing low S/N objects, we wish to remove objects with photometry affected by 
nearby neighbors. $\tt TFIT$ can identify such objects with the covariance index 
diagnostic that uses the covariance matrix \citep{Laidler:2007p2733}.
During the fitting of an object's photometry as described in \S3.4, $\tt TFIT$ uses the singular 
value decomposition routine to perform a chi-square ($\chi^2$) minimization. This yields a 
covariance matrix which is used to calculate uncertainties via the square root of the variance (the diagonal element),
as well as the covariance (the off diagonal elements) of all objects in the fit.
The covariance index is the absolute value of the ratio of the 
off-diagonal and the diagonal elements \citep{Laidler:2007p2733}. 
The maximum value of the covariance index is saved, along with the corresponding object ID, 
and yields information about how an object's photometry is affected by its most influential neighbor.
Objects for which this ratio is much less than 1 are generally isolated objects, while objects with large 
covariance index values can have unreliable photometry. 
Multiple cuts are implemented to remove objects whose photometry have been significantly affected.
First, all objects with a covariance index greater than 1 are cut because their measurements are not considered reliable.
All remaining objects are kept if one of 
two conditions apply: either they have a covariance index less than 0.5, 
or they have a $V$-band flux greater than twice that of the nearest neighbor. 
This approach balances the desire for a large sample with the need to obtain reliable photometry.
Lastly, we only consider objects that are detected
in all four ACS bands ($B$, $V$, $i^\prime$, and $z^\prime$)  to facilitate and improve color selection in \S4.2. 
This requirement removes most galaxies at $z>4$ as they have little flux in the B or redder bands. 

Table \ref{tab1} lists the 1457 galaxies  that are left after the $V$-band magnitude, covariance index, and $V$-band flux ratio cuts. 
Each entry includes the object ID (matching those from \citet{Coe:2006p1519}), R.A., decl., 
u magnitudes and uncertainties, and u S/N. It also lists information regarding the 
most significant neighbor, namely, its object ID, covariance index, $V$-band flux ratio, and separation distance.
The u magnitude uncertainties include the uncertainties due to the $V$-band aperture correction made in \citet{Coe:2006p1519},
as the accuracy of our u magnitudes depend on the accuracy of the $V$-band magnitudes.
Objects that are observed with a $u$-band flux less than 3$\sigma$ significance are considered undetected in the $u$-band
and are assigned a 3$\sigma$ upper limit.  The magnitude limit is set to:
\begin{equation}
m_{3\sigma} = -2.5 \log_{10}(3\sigma_{\mathrm{TFIT}}) + Z_{\mathrm{LRIS}},
\label{eq:maglimit}
\end{equation}
where $Z_{\mathrm{LRIS}}$ is the zero point magnitude.
The error distribution from $\tt TFIT$ is normally distributed and the upper limits from $\tt TFIT$ are robust, 
as evaluated in \citet{Laidler:2007p2733}.

\begin{deluxetable*}{ccccccccc}
\tabletypesize{\scriptsize}
\tablecaption{Catalog of u-band Objects
\label{tab1}}
\tablewidth{0pt}
\tablehead{
\colhead{ID\tablenotemark{a}} &
\colhead{R.A.} &  
\colhead{Decl.}   & 
\colhead{$u$ (mag)\tablenotemark{b}} &
\colhead{S/N}   & 
\colhead{Cov. ID\tablenotemark{c}} &
\colhead{Cov. Index\tablenotemark{d}} &
\colhead{Flux Ratio\tablenotemark{e}} & 
\colhead{Distance (arcsec)\tablenotemark{f}} }

\startdata

\vspace{0.03in}
1 &3 32 39.723 &--27 49 42.53 &24.69$ \pm $0.01 &112.53 &12 &0.039 &0.007 &2.546 \\
\vspace{0.03in}
7 &3 32 39.451 &-27 49 42.95 &25.76$ \pm $0.02 &49.56 &10 &0.413 &0.112 &1.138 \\
\vspace{0.03in}
8 &3 32 39.540 &-27 49 28.35 &26.18$ \pm $0.04 &29.43 &80 &0.003 &0.003 &4.335 \\
\vspace{0.03in}
13 &3 32 39.326 &-27 49 39.06 &26.71$ \pm $0.05 &23.76 &16 &0.095 &0.060 &2.099 \\
\vspace{0.03in}
14 &3 32 38.467 &-27 49 31.84 &25.28$ \pm $0.02 &73.11 &56 &0.010 &0.012 &3.319 \\
\vspace{0.03in}
15 &3 32 38.846 &-27 49 39.29 &28.92$ \pm $0.34 &0.34 &22 &0.159 &0.591 &1.800 \\
\vspace{0.03in}
22 &3 32 38.957 &-27 49 38.33 &28.91$ \pm $0.35 &2.25 &19 &0.281 &12.129 &1.440 \\
\vspace{0.03in}
24 &3 32 39.053 &-27 49 38.76 &27.34$ \pm $0.08 &13.27 &22 &0.296 &0.911 &1.484 \\
\vspace{0.03in}
33 &3 32 39.179 &-27 49 36.13 &27.07$ \pm $0.06 &18.38 &48 &0.018 &0.393 &2.902 \\
\vspace{0.03in}
35 &3 32 38.763 &-27 49 36.80 &27.73$ \pm $0.11 &9.78 &19 &0.026 &10.179 &2.880

\enddata

\vspace{0.03in}

\tablecomments{Table \ref{tab1} is published in its entirety in a machine-readable form in the online version of the Astrophysical Journal. A portion is shown here for guidance regarding its form and content.}
\tablenotetext{a}{ID numbers from \citet{Coe:2006p1519}.}
\tablenotetext{b}{Total $u$-band magnitudes for the same aperture as the $V$-band in \citet{Coe:2006p1519}.}
\tablenotetext{c}{ID of the most significant neighbor affecting the photometry according to the maximum covariance index.}
\tablenotetext{d}{The maximum covariance index is the absolute value of the ratio of the variance and the covariance.}
\tablenotetext{e}{The ratio of object's $V$-band flux with the $V$-band flux of its most significant neighbor.}
\tablenotetext{f}{Distance to the most significant neighbor.}

\end{deluxetable*}

\section{Photometric Selection of $z\sim3$ Galaxies}

Star-forming galaxies early in their history, such as LBGs, 
exhibit a clear break in their Spectral Energy Distribution (SED)
at the 912~\AA~Lyman limit (Lyman break), 
as well as multiple absorption lines shortward of 1216~\AA~by the Lyman series.
Photons bluer than the Lyman limit are not observed because 
the interstellar gas intrinsic to the galaxy and components of the foreground 
gas along the line of sight of the galaxy are optically thick at $\lambda \leq912$~\AA. 
At a redshift of $\gtrsim2.5$, these spectral features are redshifted to optical wavelengths, with the 
Lyman break entering the $u$-band ($\sim$3500~\AA). 
The ability to observe the Lyman break optically allows large 
samples of high-redshift galaxies to be identified based on multicolor photometry
where LBGs are selected
by a strong flux decrement shortward of the Lyman limit, and a continuum
longward of rest frame Ly$\alpha$ (1215~\AA) \citep[e.g., ][]{Steidel:1992p1911,Steidel:1995p1873, Steidel:1996p5981, Steidel:1996p5985}. 
In addition, photometric redshift determination algorithms can estimate object redshifts
using galaxy SED templates\footnote{We refer to these templates as SED templates throughout 
the paper to distinguish them from the galaxy templates discussed in \S3.},
which include additional information other than the Lyman break, such as  the slope of the SED, 
the Balmer break at 3646~\AA, and the more pronounced 4000~\AA~break, 
due to the sudden onset of stellar photospheric opacity 
by ionized metals and the CaII HK doublet \citep{Hamilton:1985p8966}.
Although color selection and photometric redshifts
utilize the same SEDs for selecting $z\sim3$ galaxies, they each have their strengths and weaknesses.
We use a combination of color selection and photometric redshifts to create our sample of LBGs (\S4.2). We
provide a description of the photometric redshift process, color selection method, 
and a catalog of all objects and their photometric redshifts below.

\begin{deluxetable*}{cccccccccccc}
\tabletypesize{\scriptsize}
\tablecaption{Catalog of Bayesian Photometric Redshifts
\label{tab2}}
\tablewidth{0pt}
\tablehead{
\colhead{ID\tablenotemark{a}} &
\colhead{$z_b$\tablenotemark{b}} &
\colhead{$t_b$\tablenotemark{c}} &
\colhead{$\tt ODDS$\tablenotemark{d}} &
\colhead{$\chi^2_\nu$\tablenotemark{e}} &
\colhead{$\chi^2_{\mathrm{mod}}$\tablenotemark{f}} & 
\colhead{$z_b1$\tablenotemark{g}} &
\colhead{$t_b1$\tablenotemark{c}} &
\colhead{$\tt ODDS1$\tablenotemark{h}} &
\colhead{$z_b2$\tablenotemark{g}} &
\colhead{$t_b2$\tablenotemark{c}} &
\colhead{$\tt ODDS2$\tablenotemark{h}} }

\startdata

\vspace{0.03in}
1 &0.17$_{-0.12}^{+0.11}$ &3.00 &1.000 &16.40 &0.20 &0.17$_{-0.07}^{+0.07}$ &3.00 &1.000 &\nodata &\nodata &\nodata \\
\vspace{0.03in}
7 &0.01$_{-0.01}^{+0.10}$ &6.67 &0.990 &7.59 &0.49 &0.01$_{-0.01}^{+0.10}$ &6.67 &0.990 &1.81$_{-0.02}^{+0.02}$ &6.00 &1.000 \\
\vspace{0.03in}
8 &0.52$_{-0.15}^{+0.15}$ &1.00 &1.000 &21.81 &0.01 &0.52$_{-0.08}^{+0.07}$ &1.00 &1.000 &\nodata &\nodata &\nodata \\
\vspace{0.03in}
13 &0.43$_{-0.14}^{+0.14}$ &3.33 &1.000 &8.00 &0.21 &0.43$_{-0.08}^{+0.11}$ &3.33 &1.000 &\nodata &\nodata &\nodata \\
\vspace{0.03in}
14 &0.54$_{-0.15}^{+0.15}$ &2.33 &1.000 &46.40 &0.06 &0.54$_{-0.08}^{+0.07}$ &2.33 &1.000 &\nodata &\nodata &\nodata \\
\vspace{0.03in}
15 &0.52$_{-0.16}^{+0.15}$ &1.67 &0.970 &1.98 &0.09 &0.52$_{-0.12}^{+0.03}$ &1.67 &0.520 &0.57$_{-0.02}^{+0.10}$ &2.00 &0.443 \\
\vspace{0.03in}
22 &3.23$_{-0.41}^{+0.41}$ &5.00 &0.999 &7.34 &0.44 &3.23$_{-0.29}^{+0.19}$ &5.00 &0.977 &2.90$_{-0.10}^{+0.04}$ &3.33 &0.023 \\
\vspace{0.03in}
24 &1.39$_{-0.23}^{+0.78}$ &7.00 &0.363 &1.56 &1.68 &1.39$_{-0.13}^{+0.12}$ &7.00 &0.247 &1.65$_{-0.14}^{+0.22}$ &6.67 &0.429 \\
\vspace{0.03in}
33 &0.78$_{-0.17}^{+0.17}$ &3.67 &1.000 &1.96 &0.08 &0.78$_{-0.09}^{+0.09}$ &3.67 &1.000 &\nodata &\nodata &\nodata \\
\vspace{0.03in}
35 &1.69$_{-0.26}^{+0.26}$ &6.00 &0.996 &4.81 &0.60 &1.69$_{-0.27}^{+0.16}$ &6.00 &0.996 &0.02$_{-0.01}^{+0.03}$ &6.67 &0.004 

\enddata

\vspace{0.03in}

\tablecomments{Table \ref{tab2} is published in its entirety in a machine-readable form in the online version of the Astrophysical Journal. A portion is shown here for guidance regarding its form and content. In addition to the most likely redshift, we report the two most likely redshifts for each galaxy when available, along with the redshift ranges for each peak and the fractions of $P(z)$ contained in those peaks. }
\tablenotetext{a}{ID numbers from \citet{Coe:2006p1519}.}
\tablenotetext{b}{Bayesian photometric redshift (BPZ)  and uncertainty from 95\% confidence interval.}
\tablenotetext{c}{Template SEDs used in the BPZ code as described in \S4.1.1, where 1=El\_cww, 2=Scd\_cww, 3=Sbc\_cww, 4=Im\_cww, 5=SB3\_kin, 6=SB2\_kin , 7=25Myr, and 8=5Myr. Non-interger values are for templates interpolated between adjacent templates.}
\tablenotetext{d}{Integrated $P(z)$ contained within $0.1(1+z_b)$.}
\tablenotetext{e}{Chi square quality of photo-$z$ fit.}
\tablenotetext{f}{Modified reduced chi-square fit, where the templates are given uncertainties.}
\tablenotetext{g}{Redshift ranges for the two most likely peaks.}
\tablenotetext{h}{Integrated $P(z)$ contained within the local minima of $P(z)$ for each peak.}

\end{deluxetable*}


\subsection{Photometric Redshifts}

Photometric redshifts (hereafter, photo-$z$'s) are a well known and robust procedure to determine 
redshifts of galaxies when spectra are unavailable 
\citep[e.g.][]{Koo:1985p9108, Lanzetta:1998p9100, Benitez:2000p3572, Coe:2006p1519, Hildebrandt:2008p2281}.
They have the advantage over color 
selection that they take into account all the colors available simultaneously in $\chi^2$ fits to template SEDs
and yield more precise redshift information with clear redshift confidence limits. 
The photo-$z$'s also sample $z\sim3$ galaxies in regions of color space that
color selected samples avoid because of low-redshift galaxies, and therefore can provide a larger
sample. However, photo-$z$ uncertainties do not always include systematic errors caused by variations and evolution of 
galaxy SEDs compared to SED templates, and possible mismatches of SED templates (see \S4.1.2). Such systematic
problems and the lack of a large spectroscopic sample make it difficult to characterize the contamination fraction of 
photo-$z$ selected $z\sim3$ galaxies (see \S4.1.3). Nonetheless, they provide the largest sample of LBGs for study.

For each galaxy, photo-$z$ codes produce a probability distribution function, $P(z)$, 
representing the probability of a galaxy being at any specific redshift. 
However, the $P(z)$ can have multiple peaks, especially 
at $z\sim3$ where there is a degeneracy 
with galaxies at $z\sim0.2$, which then translates into 
large uncertainties for the photo-$z$ \citep{Benitez:2000p3572}.
The introduction of the $u$-band data helps resolve the photo-$z$ degeneracy 
and improve the photo-$z$ fits for $z\sim3$ galaxies
as it targets the most dominant signature in their SED, the Lyman break.
We present photo-$z$'s for the entire sample of galaxies with $u$-band data from \S3.5 in Table \ref{tab2},
but caution against using them blindly to select galaxies at $z\sim3$. We recommend 
making cuts on the sample to select galaxies with good $\chi^2_{\mathrm{mod}}$ and $\tt ODDS$ (for a description of these parameters, see \S4.1.1).

\subsubsection{Bayesian Photometric Redshifts}

There are many different photo-$z$ codes available, and 
\citet{Hildebrandt:2008p2281} explain the benefits of the different algorithms. 
We chose to use the Bayesian photo-$z$'s (BPZ) \citep{Benitez:2000p3572, Benitez:2004p3578, Coe:2006p1519}
to be consistent with past photo-$z$'s determined for the UDF without $u$-band data \citep{Coe:2006p1519}. 
\citet{Hildebrandt:2008p2281} advise using the SED templates supplied with their respective codes, 
since user-supplied SED templates can cause problems.
However, a re-calibration of the SED template set improves the performance of the photo-$z$ redshifts, and we use
the re-calibrated SED templates from \citet{Benitez:2004p3578} and \citet{Coe:2006p1519} that have been extensively tested with BPZ.
These re-calibrated SEDs are based on the star-forming (Im), spiral (Scd, Sbc), 
and elliptical (El) galaxy templates from \citet{Coleman:1980p4084},  the star bursting galaxy templates
 with different reddening (SB2, SB3) from
\citet{Kinney:1996p6459}, and the faint blue galaxy SEDs with ages of 25 and 5 Myr and 
metallicities of Z=0.08 without dust from \citet{Bruzual:2003p4897}, 
described in section \S4.1 of \citet{Coe:2006p1519}. 
We interpolate between adjacent galaxy SED templates for an additional two SED templates in the photo-$z$ fit, similar to
\citet{Benitez:2004p3578} and \citet{Coe:2006p1519}.

SED templates are not always a good match for each specific galaxy, and when it is not possible to
get a good fit to the SED template, then the resulting redshift may not be accurate. As a diagnostic
of the goodness of fit, the BPZ code provides a reduced chi square ($\chi^2_\nu$) value. However, high $\chi^2_\nu$ values
do not always indicate an unreliable redshift. Bright galaxies with small photometric uncertainties will have larger 
$\chi^2_\nu$ values than faint galaxies with larger photometric uncertainties for the same numerator 
in $\chi^2$ (also known as the variance), yet have more reliable redshifts \citep[see Figure 21 in][]{Coe:2006p1519}.
This problem occurs because the systematic uncertainties of the SED templates are not taken into account, making
$\chi^2_\nu$ no longer represent the relative quality of the fit. 
Nonetheless, a mechanism to evaluate the quality of the fits is required to trim the sample to reliable redshifts. 
To this end, \citet{Coe:2006p1519} introduce a 
modified reduced chi square ($\chi^2_{\mathrm{mod}}$) value that assigns an uncertainty to the SED templates in addition
to the uncertainty in the photometry of the galaxy.
For clarity, we reproduce the equation from \citet{Coe:2006p1519} here:
\begin{equation}
\chi^2_{\mathrm{mod}}=\sum_{\alpha}\frac{(f_\alpha-f_{T\alpha})^2}{\sigma^2_{f_\alpha}+\sigma^2_{f_T}}/\nu, 
\label{eq:chi}
\end{equation}
where $f_\alpha$ are the observed fluxes, $\sigma_{f_\alpha}$ is the error in observed fluxes, 
and $f_{T_\alpha}$ are the model fluxes, normalized to the observed fluxes.
$\sigma_{f_{T_\alpha}}$  represent the model flux errors, which are set by \citet{Coe:2006p1519} 
to $\sigma_{f_{T_\alpha}}=$max$_\alpha(f_{T\alpha})/15$. While the definition of $\sigma_{f_{T_\alpha}}$ is
arbitrary, it was picked such that the resultant $\chi^2_{\mathrm{mod}}$ is a more realistic measure of the goodness of fit. 
This is especially important for bright galaxies, as uncertainties in the templates
dominate the error budget, and the $\chi^2_{\nu}$ values are not useful. The reported $\chi^2_{mod}$ values are reduced
chi square values, obtained by dividing by the number of degrees of freedom, $\nu$.  The number of degrees of 
freedom is the difference between the number of filters observed and 
the number of parameters (in this case there are three fit parameters, redshift, template, and amplitude).
The minimum number of filters used is 5 and the maximum is 7, so the range of $\nu$ in our study is $2\leq \nu \leq4$. 
We note that $\chi^2_{\mathrm{mod}}$ is calculated after the photo-$z$ determinations, and does not affect $P(z)$. 

If the quality of the fit to the SED template is good, then the $\tt ODDS$ parameter is useful in measuring the spread in $P(z)$. 
A galaxy with high $\tt ODDS$ has a single peak in $P(z)$, while multiple or very wide peaks yield low $\tt ODDS$. 
In general, restricting the photometric sample to those objects with $\tt ODDS$ $> 0.9--0.99$
yield clearly defined redshifts \citep{Benitez:2000p3572, Benitez:2004p3578, Coe:2006p1519}. 
In this paper we are conservative and restrict our sample to objects with the best vales of $\tt ODDS$, those with $\tt ODDS$ $>0.99$.
Additionally, selecting galaxies based on SED template type ($t_b$) can useful when selecting a specific type of galaxy,
where 1=El\_cww, 2=Scd\_cww, 3=Sbc\_cww, 4=Im\_cww, 5=SB3\_kin, 6=SB2\_kin , 7=25Myr, and 8=5Myr. 
For instance, in selecting LBGs we constrain ourselves to galaxies with $t_b>3$, which only include star-forming galaxy templates.
The redshifts, redshift uncertainties for a 95\% confidence interval, $t_b$, $\tt ODDS$, $\chi^2_\nu$, and  $\chi^2_{\mathrm{mod}}$ are all tabulated in Table \ref{tab2}.

\vspace{0.2in}

\subsubsection{Photometric Redshift Measurement Uncertainties}

It is important to understand the origins of photo-$z$ uncertainties in order to have confidence in their values.
There are two types of uncertainties in photo-$z$'s: 1) photometric measurement uncertainties and 2) template mismatch variance
\citep{Lanzetta:1998p9100,FernandezSoto:2001p2773, FernandezSoto:2002p9103, Chen:2003p8912}.
Photometric measurement uncertainties are well understood, and they are responsible for the width of $P(z)$, 
which determines the reported photo-$z$ uncertainties. Faint galaxies with larger photometric measurement uncertainties will yield 
larger uncertainties in the photo-$z$'s than brighter galaxies. 
If the photometric measurement uncertainties are very large, then the photo-$z$ will be poorly constrained, 
because this results in multiple peaks in $P(z)$ corresponding to many possible redshifts. 
The other possible uncertainty comes from template mismatches, which is a systematic error
due to the finite number of templates used in the photo-$z$ determination. 
Not all galaxies will be well represented by our template SEDs, yielding large $\chi^2_{\nu}$ values. 

One method to decrease template mismatch errors is to introduce more template SEDs, since that increases the chance that
there exist good matching SED templates for each galaxy. 
While this can improve low-redshift performance, it also increases the number of degenerate 
solutions and therefore gives poorer high-redshift performance \citep{Hildebrandt:2008p2281}.
Degenerate photo-$z$'s occur when different SED templates fit the photometry equally well, 
resulting in multiple peaks in $P(z)$ and therefore multiple possible redshifts.
We consider galaxy redshifts degenerate if they have two or more peaks in $P(z)$ at 95\% confidence separated by $\Delta z > 1$. 
In these cases, the resultant reported uncertainties are very large and the galaxy redshift is poorly constrained. 
We are mainly interested in good performance at high redshift and therefore 
do not increase the number of SED templates. 

Template mismatches are also the cause for ``catastrophic" photo-$z$ errors that occur, where the photo-$z$ is incorrect
and the uncertainty does not include the correct redshift \citep{Ellis:1997p3771, FernandezSoto:1999p2784, Benitez:2000p3572}. 
Catastrophic photo-$z$ errors typically occur because multiple peaks in $P(z)$ are incorrectly suppressed, leaving only one peak. 
The suppression of multiple peaks in bright galaxies likely occur because of their small photometric 
uncertainties yielding large $\chi^2_\nu$ values without taking into account the systematic uncertainties in the SED templates, 
which exaggerate the differences between the different SED template fits that correspond to different peaks in $P(z)$.
This suppresses the peaks at other redshifts, resulting in possible ``catastrophic" photo-$z$ errors (Dan Coe, private communication).

It is likely that the incorrect suppression of peaks in $P(z)$ can be fixed through the introduction of SED uncertainties in the initial $\chi^2$ fit,
although such an addition requires an understanding of those uncertainties using large surveys with both photo-$z$ and 
spec-$z$'s. This is neither in the scope of the UDF or this paper, and is being investigated elsewhere (Coe et al. in prep). 
In the mean time, the best we can do is reject photo-$z$'s based on their $\chi^2_{\mathrm{mod}}$ values.
However, $\chi^2_{\mathrm{mod}}$ is not a true statistical test like $\chi^2_{\nu}$, and therefore cuts normally appropriate for
$\chi^2_{\nu}$ are not valid for $\chi^2_{\mathrm{mod}}$. Additionally, galaxies with $J$ and $H$-bands data have median $\chi^2_{\mathrm{mod}}$ larger
by $\sim0.7$ than those without infrared (IR) data, although the inclusion of the IR data improves the reliability of the photo-$z$'s 
because of the significantly increased lever arm \citep{Coe:2006p1519}. We therefore don't use the same cut as used in 
\citet{Coe:2006p1519} ($\chi^2_{\mathrm{mod}} < 1$), but rather use $\chi^2_{\mathrm{mod}}$ to conservatively remove possible bad photo-$z$ fits.
While we do present all the photo-$z$ fits, we only include those with $\chi^2_{\mathrm{mod}} < 4$ for our $z\sim3$ galaxy sample. 
This cut only reduces the total number of galaxies with photo-$z$'s by $\sim$5\% (to 1385 galaxies).

\begin{figure*}
\plottwo{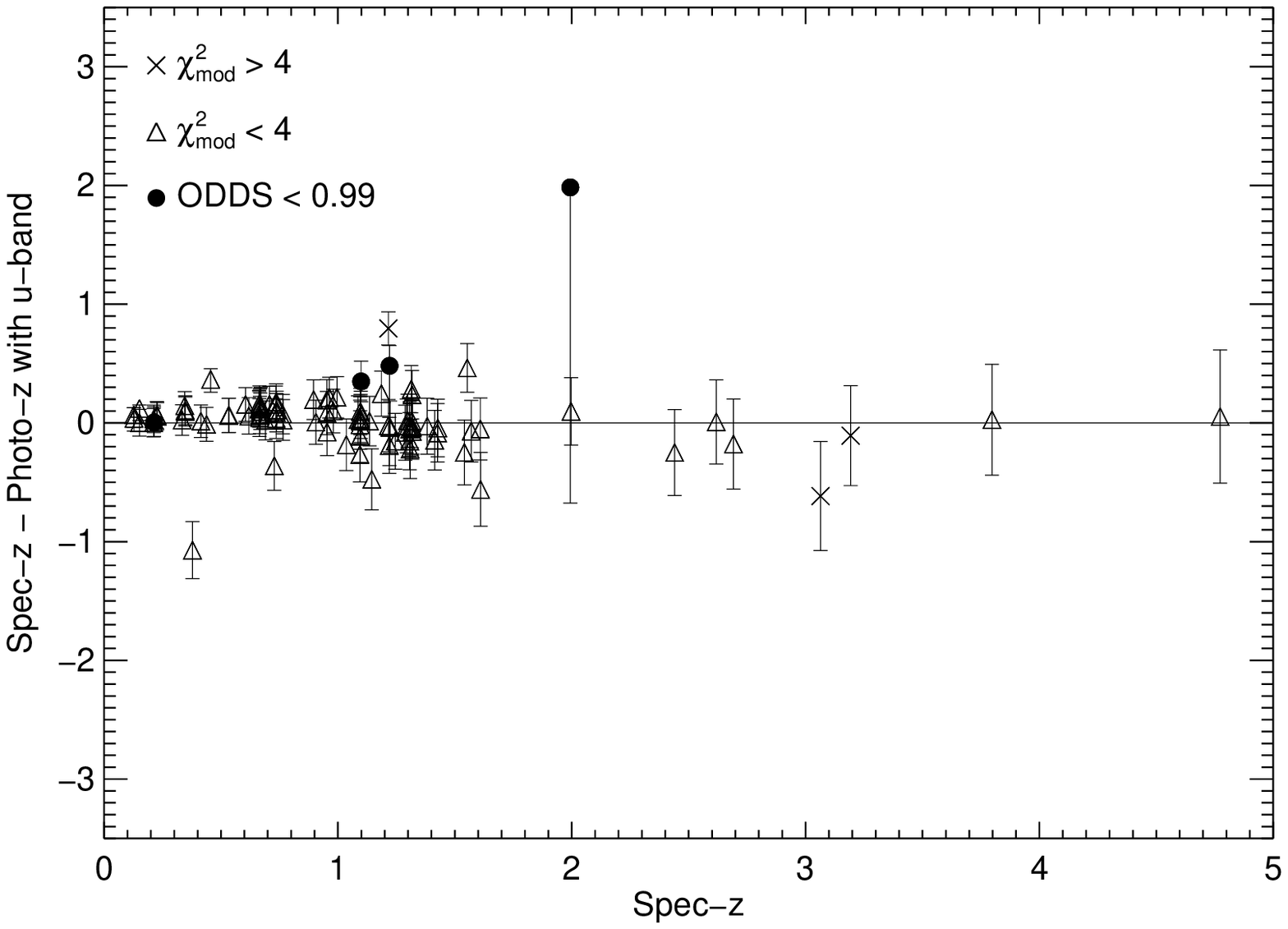}{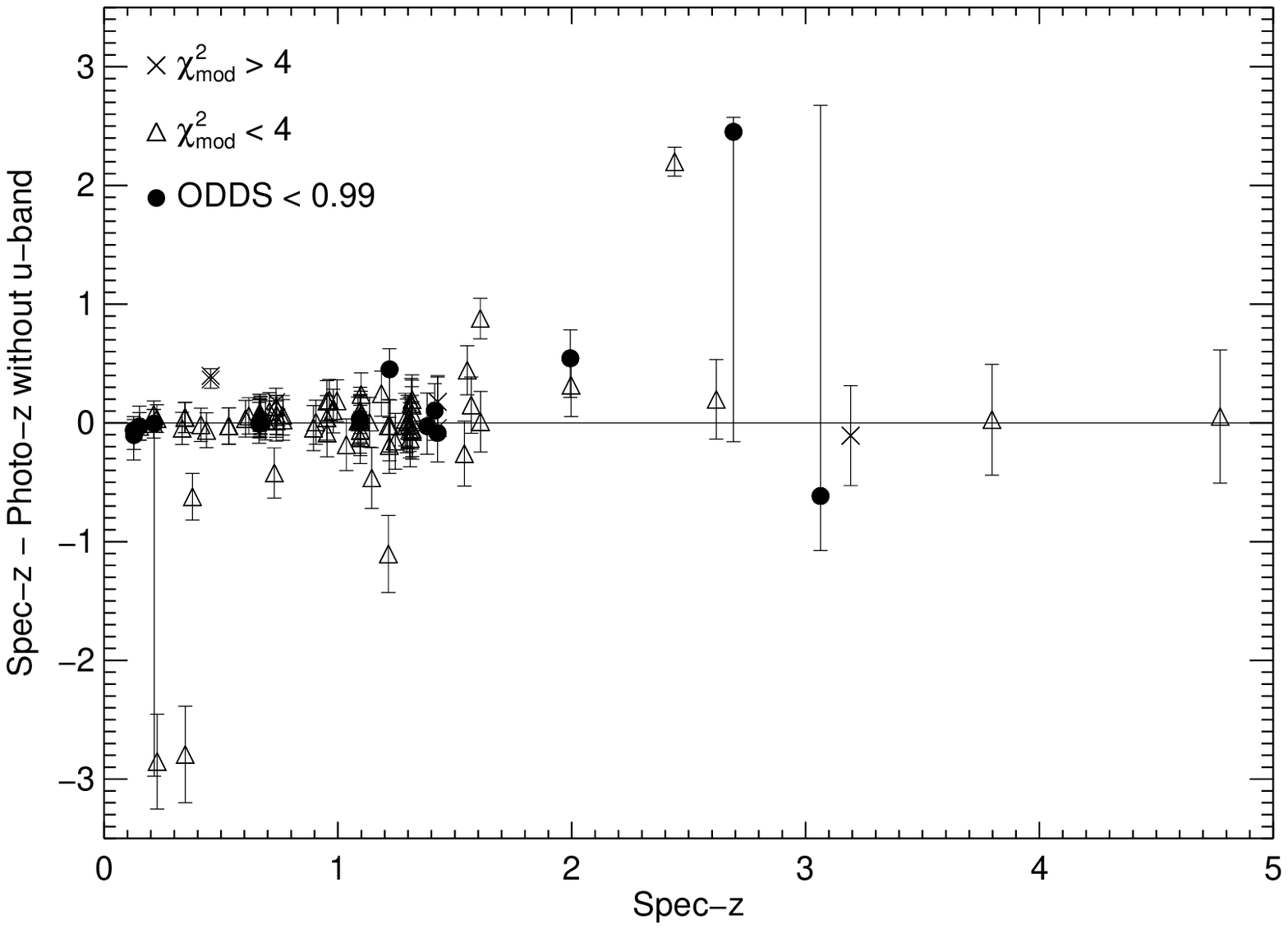}
\caption{Differences in spectroscopic redshifts (spec-$z$'s) and photometric redshifts (photo-$z$'s) compared to the 100 spec-$z$'s described in \S4.1.3, 
where the  photo-$z$ uncertainties are for a 95\% confidence interval. 
The left panel plots the difference in spec-$z$ and photo-$z$ with $u$-band, and the right panel plots the same but for the 
photo-$z$'s without the  $u$-band. 
The quality of the photometric redshift fit is described by the $\chi^2_{\mathrm{mod}}$ (see \S4.1.1), where
objects with good fits, $\chi^2_{\mathrm{mod}} < 4$,  are marked with triangles, and objects with
$\chi^2_{\mathrm{mod}} > 4$ are marked with open crosses. 
There is one point in the left panel at $z\sim$2 that has a large $\chi^2_{\mathrm{mod}}$ ($\sim17$) and should be ignored as it is not a good fit to any SED templates with the $u$-band.
\label{photzspec}}
\end{figure*}

\begin{deluxetable*}{crcccccccc}
\tabletypesize{\scriptsize}
\tablecaption{Catalog of Galaxies with Reliable Spectroscopic Redshifts in the UDF
\label{tab3}}
\tablewidth{0pt}
\tablehead{
\colhead{ID\tablenotemark{a}} &
\colhead{Survey} &
\colhead{$z_{\mathrm{spec}}$} &
\colhead{$z_b$\tablenotemark{b}} &
\colhead{$\chi^2_{\mathrm{mod}}$\tablenotemark{c}} & 
\colhead{$\tt ODDS$\tablenotemark{d}} &
\colhead{V (mag)} &
\colhead{$u-V$ (mag)} &
\colhead{$u-B$ (mag)} &
\colhead{$V-z^\prime$ (mag)}}

\startdata

\vspace{0.03in}
3088 &FORS2 &0.13 &0.06$_{-0.06}^{+0.10}$ &1.91 &1.00 &22.85$ \pm $0.00 &1.99$ \pm $0.01 &1.16$ \pm $0.01 &0.48$ \pm $0.00 \\
\vspace{0.03in}
5670 &VVDS &0.13 &0.09$_{-0.09}^{+0.11}$ &0.03 &1.00 &21.23$ \pm $0.00 &2.00$ \pm $0.00 &1.12$ \pm $0.00 &0.54$ \pm $0.00 \\
\vspace{0.03in}
1971 &VVDS &0.15 &0.03$_{-0.03}^{+0.10}$ &0.07 &1.00 &20.46$ \pm $0.00 &1.48$ \pm $0.00 &0.82$ \pm $0.00 &0.37$ \pm $0.00 \\
\vspace{0.03in}
2974 &VIMOS &0.15 &0.15$_{-0.11}^{+0.11}$ &1.51 &1.00 &24.08$ \pm $0.01 &1.97$ \pm $0.03 &0.99$ \pm $0.03 &0.49$ \pm $0.01 \\
\vspace{0.03in}
5620 &VVDS &0.21 &0.21$_{-0.12}^{+0.12}$ &0.75 &1.00 &23.42$ \pm $0.00 &0.88$ \pm $0.01 &0.43$ \pm $0.01 &0.01$ \pm $0.00 \\
\vspace{0.03in}
3822 &VIMOS &0.21 &0.18$_{-0.12}^{+0.12}$ &0.59 &1.00 &19.14$ \pm $0.00 &1.91$ \pm $0.00 &0.87$ \pm $0.00 &0.70$ \pm $0.00 \\
\vspace{0.03in}
1000 &FORS2 &0.21 &0.21$_{-0.13}^{+0.12}$ &1.08 &0.97 &23.39$ \pm $0.00 &0.93$ \pm $0.01 &0.46$ \pm $0.01 &-0.01$ \pm $0.01 \\
\vspace{0.03in}
5606 &VVDS &0.23 &0.16$_{-0.11}^{+0.11}$ &0.05 &1.00 &21.14$ \pm $0.00 &1.79$ \pm $0.00 &0.82$ \pm $0.00 &0.55$ \pm $0.00 \\
\vspace{0.03in}
5491 &VIMOS &0.23 &0.17$_{-0.12}^{+0.11}$ &1.94 &1.00 &22.24$ \pm $0.00 &0.64$ \pm $0.00 &0.30$ \pm $0.00 &-0.15$ \pm $0.00 \\
\vspace{0.03in}
7847 &VVDS &0.33 &0.31$_{-0.13}^{+0.13}$ &0.02 &1.00 &22.00$ \pm $0.00 &2.97$ \pm $0.01 &1.36$ \pm $0.01 &1.10$ \pm $0.00

\enddata

\vspace{0.03in}

\tablecomments{Table \ref{tab3} is published in its entirety in a machine-readable form in the online version of the Astrophysical Journal. 
A portion is shown here for guidance regarding its form and content. Redshift surveys are VVDS \citep{LeFevre:2004p3988}, 
VIMOS \citep{Popesso:2009p6629}, \citet{Szokoly:2004p4004}, and 
FORS2 \citep{Vanzella:2005p6608, Vanzella:2006p6605, Vanzella:2008p3704, Vanzella:2009p9955}. V magnitudes are
total AB magnitudes, and colors are isophotal colors. All photometry other than the $u$-band are from \citet{Coe:2006p1519}.
Nondetections in u-band are given 3$\sigma$ limiting magnitudes. }
\tablenotetext{a}{ID numbers from \citet{Coe:2006p1519}.}
\tablenotetext{b}{Bayesian Photometric Redshift (BPZ)  and uncertainty from 95\% confidence interval.}
\tablenotetext{c}{Modified reduced chi square fit, where the templates are given uncertainties.}
\tablenotetext{d}{Integrated $P(z)$ contained within $0.1(1+z_b)$.}

\end{deluxetable*}

\begin{figure*}
\plottwo{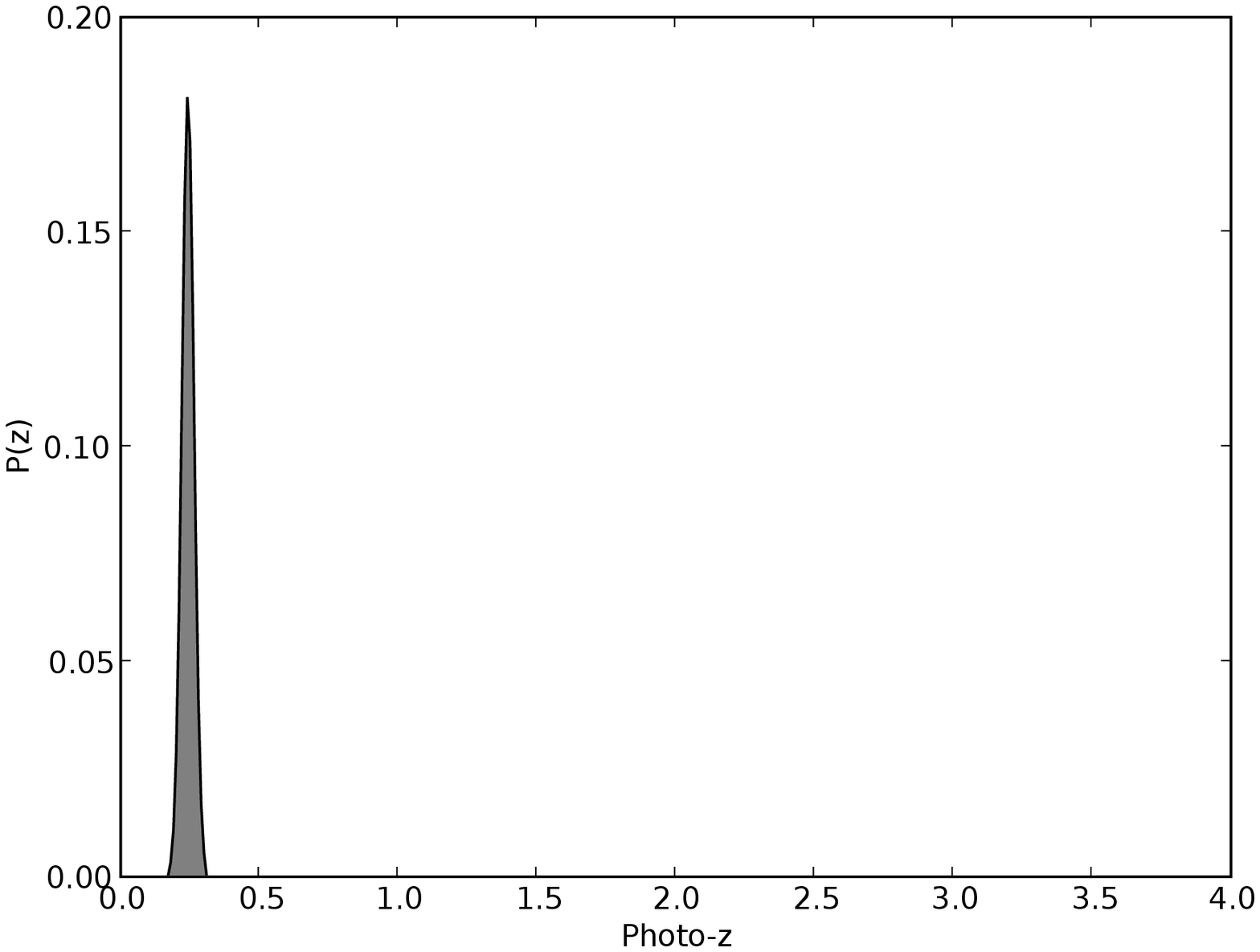}{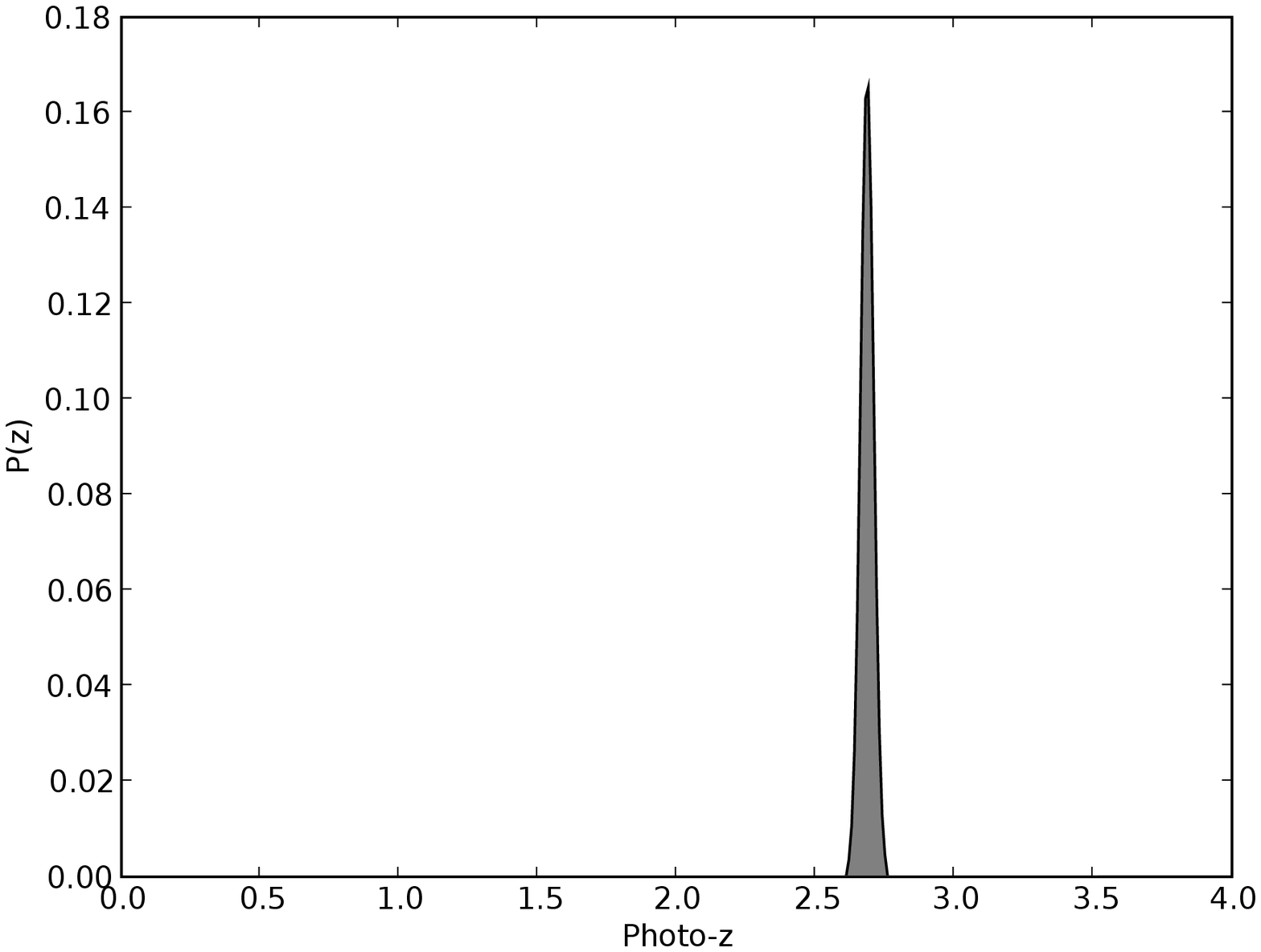}
\plottwo{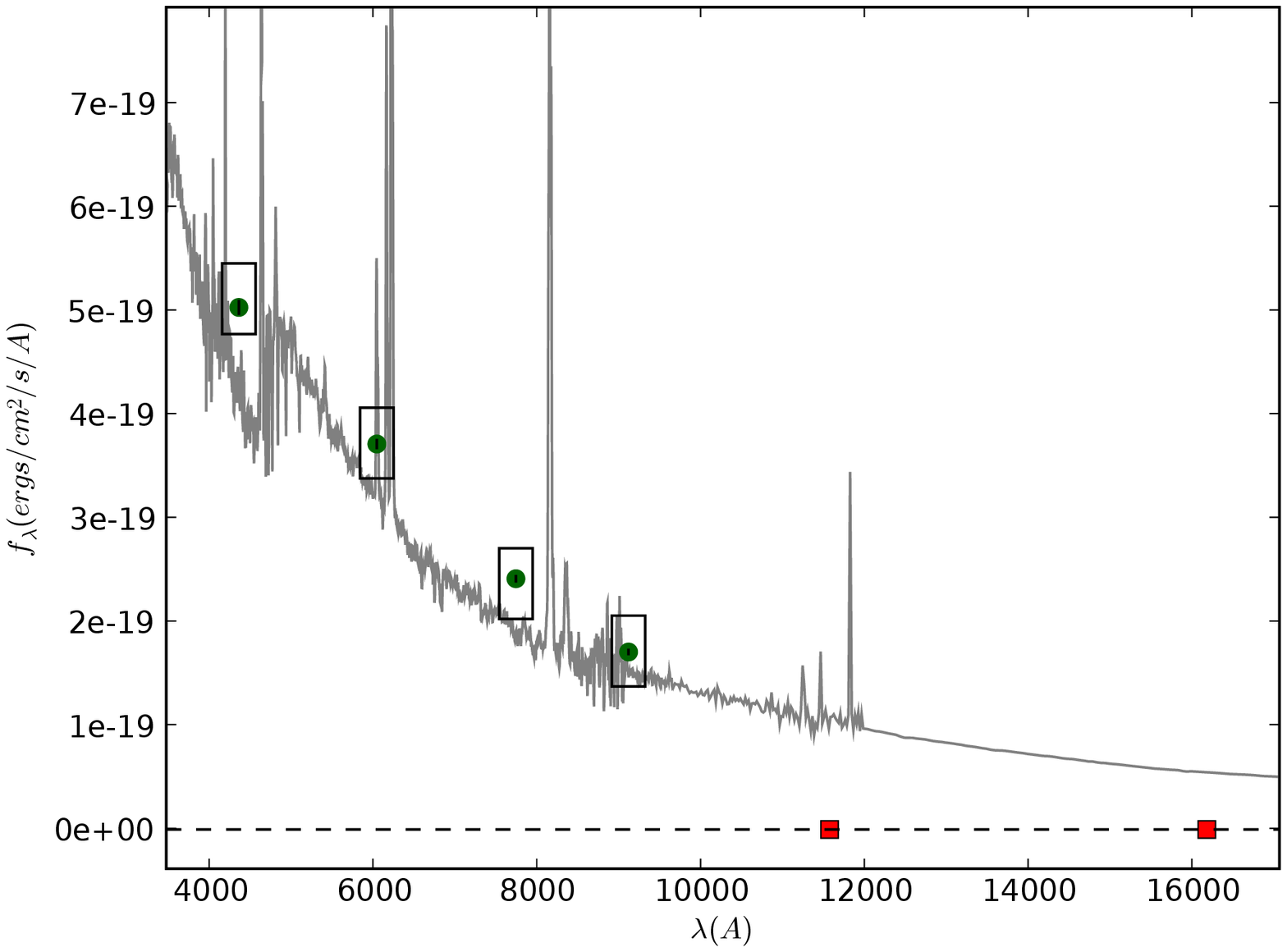}{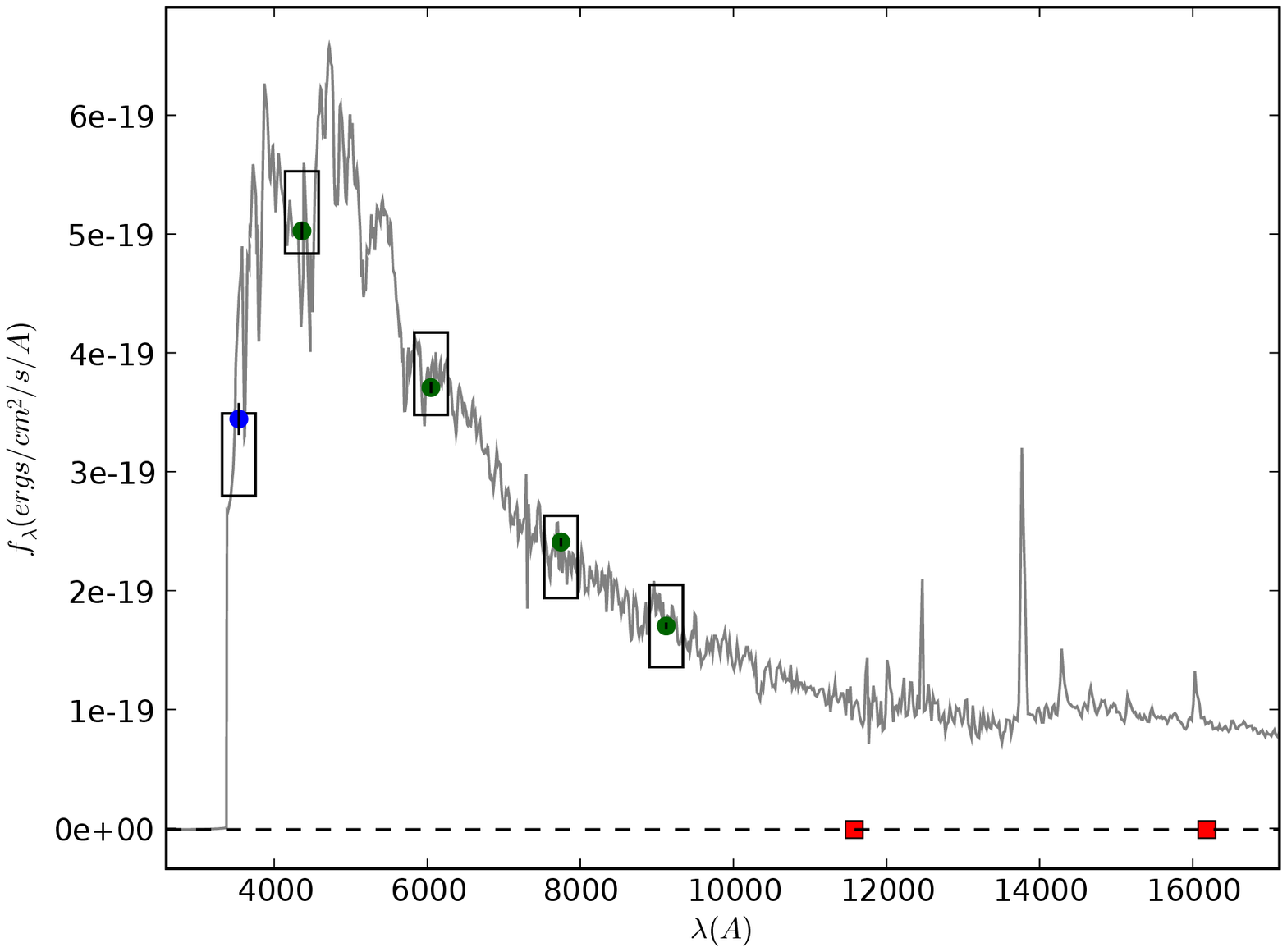}
\caption{Object 830: an example of a galaxy with $z_{\mathrm{spec}} = 2.44$ \citep{Popesso:2009p6629} 
that has a catastrophic redshift error before the addition of the $u$-band. The left-hand side figures show the 
photo-$z$ fit without $u$-band with z=0.24$\pm0.15$ and the right-hand side figures show the fit with $u$-band with z=2.72$\pm0.37$.
The rectangles represent the assumed uncertainty of the SED template used for $\chi^2_{\mathrm{mod}}$, blue points are the $u$-band 
data, green are the ACS data, and red are the NICMOS data. Filled circles are for detected and observed data points, and squares are for 
unobserved data points. 
The addition of the $u$-band rules out the possibility of this being a low-redshift galaxy, and corrects the 
catastrophic redshift.
\label{obj830}}
\end{figure*}

\subsubsection{Comparison of Photometric and Spectroscopic Redshifts}

In order to test the accuracy of the photo-$z$'s, we compare the redshifts with
spectroscopic redshifts (spec-$z$'s). We compile a list of 100 
reliable spec-$z$'s in the UDF that match our sample from \S3.5
(see Table \ref{tab3})\footnote{
Most of these redshifts are based on observations made with ESO Telescopes 
at the La Silla or Paranal Observatories under programme ID(s) 
66.A-0270(A), 67.A-0418(A),171.A-3045, 170.A-0788, 074.A-0709, and 275.A-5060.}. 
In this sample, 18 spec-$z$'s are from
the VIMOS VLT Deep Survey \citep[VVDS;][]{LeFevre:2004p3988}, 
where we only include redshifts with 95\% confidence and multiple lines.
Another 22 redshifts come from the GOODS VLT VIMOS survey \citep[VIMOS;][]{Popesso:2009p6629}, 
where we include redshifts with A or B quality spectra. These spectra have good cross-correlation coefficients of the spectra with the templates and 
multiple lines are well identified. 
An additional 6 redshifts are from \citet{Szokoly:2004p4004} using the VLT FORS1/FORS2 spectrographs, 
where we use only those flagged as 'reliable' redshifts (quality flags ``2" or ``2+").
The remaining 57 redshifts are from
the GOODS VLT FORS2 survey 
\citep[FORS2;][]{Vanzella:2005p6608, Vanzella:2006p6605, Vanzella:2008p3704, Vanzella:2009p9955},
where we only include redshifts from A or B quality spectra. 
We do not include redshifts from the slitless spectra obtained as part of the Grism ACS Program for Extragalactic Science (GRAPES) \citep{Pirzkal:2004p9474}, 
since the redshift determinations do not provide an independent check to photo-$z$'s because photo-$z$'s were used to help identify the emission lines \citep{Xu:2007p4282}.

The photo-$z$'s agree relatively well with the spec-$z$'s (see left panel of Figure \ref{photzspec}), and have 100\% agreement
in the redshift interval of interest ($2.5\lesssim z \lesssim 3.5$), although only five objects have spec-$z$'s at these redshifts. 
There are clearly some galaxies that have incorrect photo-$z$'s at lower 
redshift where the $u$-band does not sample the Lyman break. 
Of the 100 galaxies from Table \ref{tab3}, 97 have $\chi^2_{\mathrm{mod}} < 4$, of which 93 have $\tt ODDS$ $ > 0.99$. 
The galaxy at a spec-$z$ of 1.99 (ID 6834) has a $\chi^2_{\mathrm{mod}} \sim17$ and the resultant photo-$z$ should be ignored.

Only object 8585 meets our criteria, but has a significantly different photo-$z$ than its spec-$z$ and does not include the correct redshift in its $P(z)$. 
This object, with a spec-$z$ of $z=0.3775$ \citep{LeFevre:2004p3988} and a photo-$z$ of $1.45\pm0.24$, is a case where the spec-$z$ may be wrong. 
In general we tried to minimize this possibility by selecting reliable spec-$z$'s, but the spec-$z$'s can still be wrong as seen in some comparisons of \citet{FernandezSoto:2001p2773}.
In our case, the photometric redshift yields a good fit to the SED templates,  
and the fit to the SED template at the spec-$z$ is a bad fit. The published spectrum shows that the spec-$z$ is mainly determined by the $H\alpha$ line.
This could easily be confused with the O$\rm{II}$ line for a galaxy with a redshift of $z=1.43$, which would be consistent with our photo-$z$.
This leaves one galaxy with a possible ``catastrophic error", although
it is not in the redshift interval of interest ($2.5\lesssim z \lesssim 3.5$).

\begin{figure}
\plotone{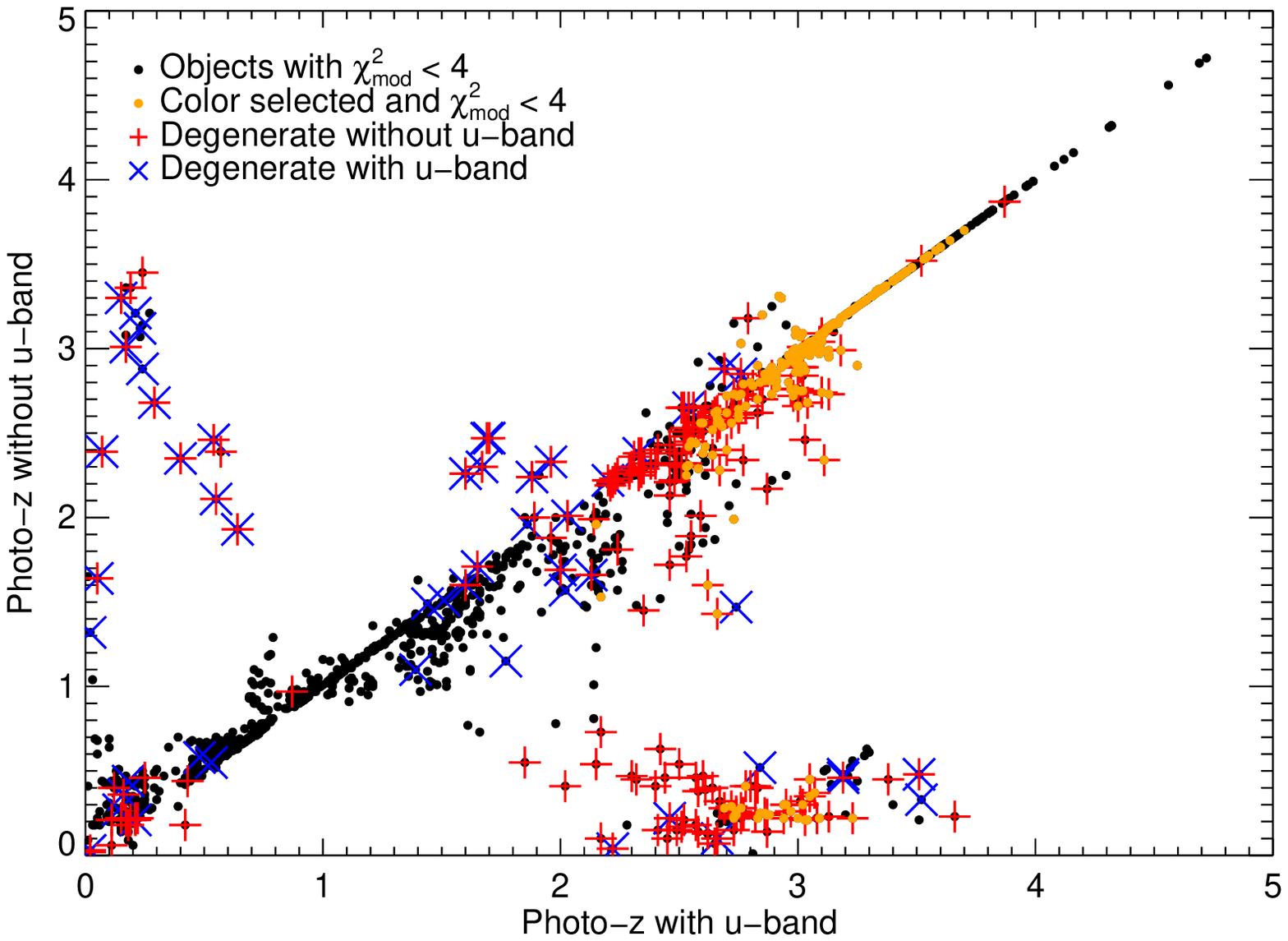}
\caption{Photo-$z$'s without $u$-band data versus photo-$z$'s with $u$-band data for objects with $\chi^2_{\mathrm{mod}} < 4$ in both cases.
The red crosses designate galaxies where the photo-$z$'s are degenerate without $u$-band, and the blue crosses 
are for degenerate photo-$z$'s with $u$-band. 
The degeneracy in redshift occurs when galaxies have at least two significant peaks in the $P(z)$, usually due to the degeneracy 
in the colors of a low-redshift galaxies and high-redshift galaxies. 
The orange dots are those objects that are selected using the color selection technique in \S4.2. 
We note although the photo-$z$'s with $u$-band may be incorrect, the addition of the $u$-band
made a large difference in the photo-$z$'s of $z\sim3$ galaxies. 
\vspace{0.2in}
\label{photzcomp}}
\end{figure}

\begin{figure*}
\plottwo{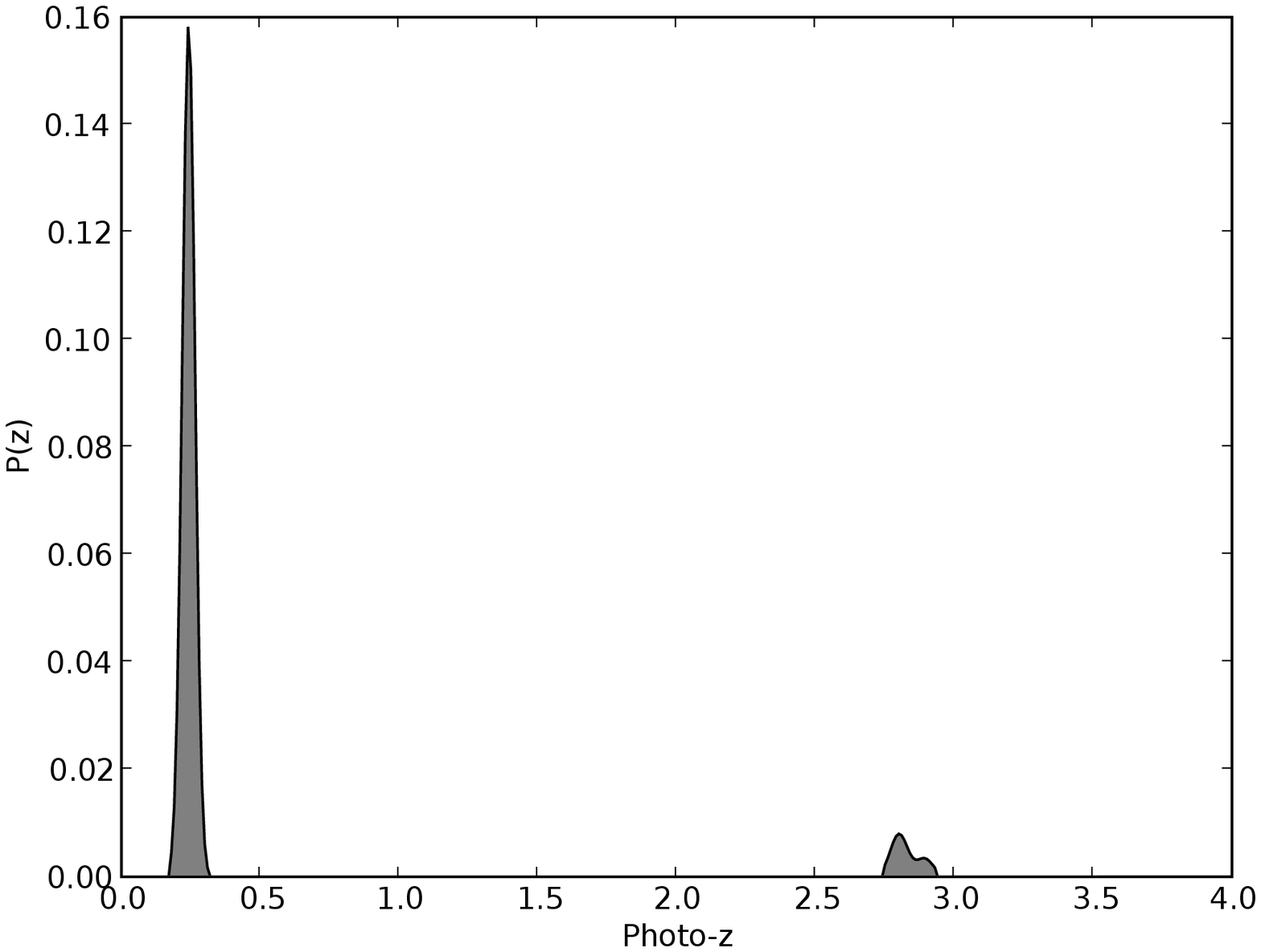}{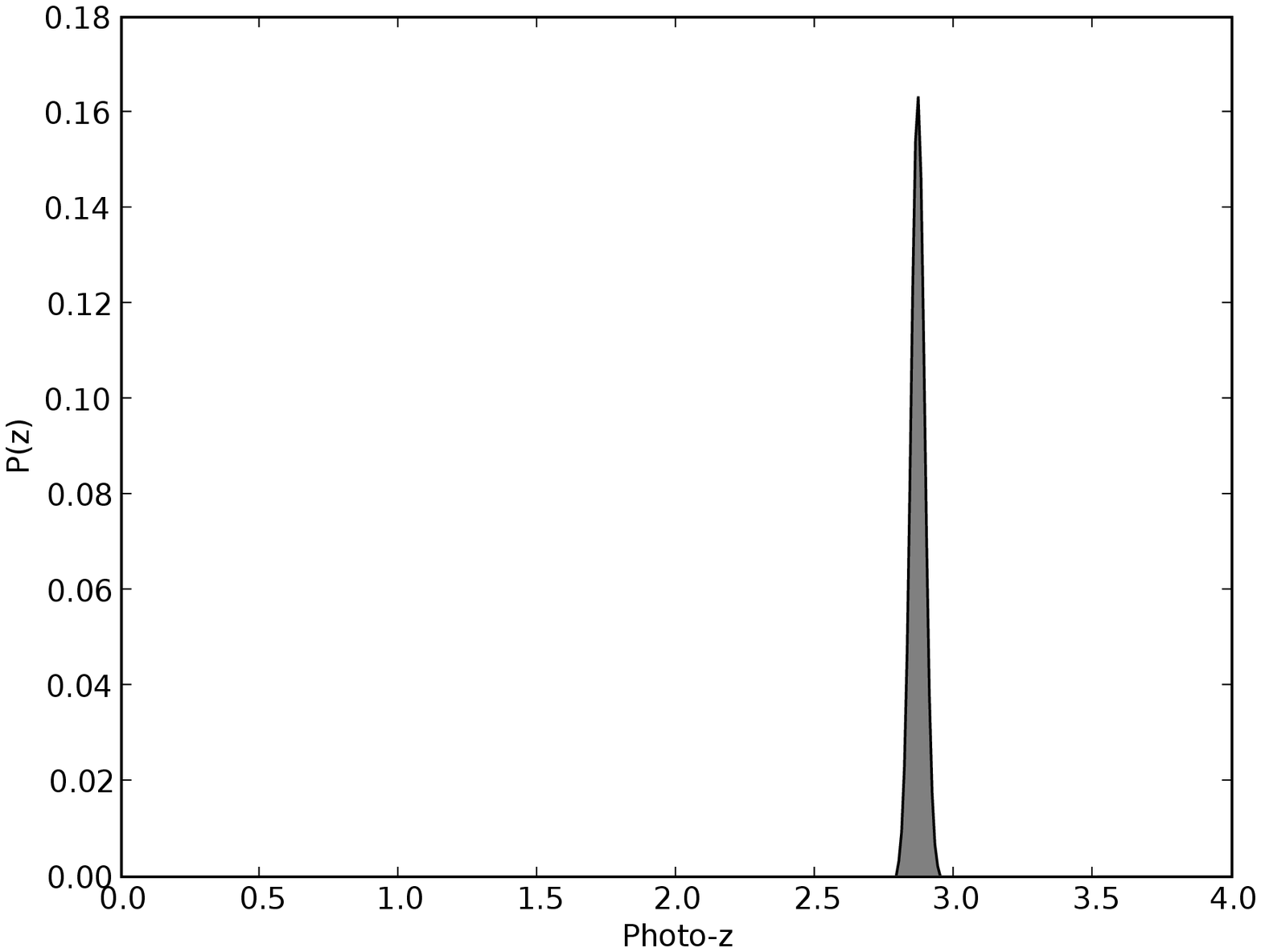}
\plottwo{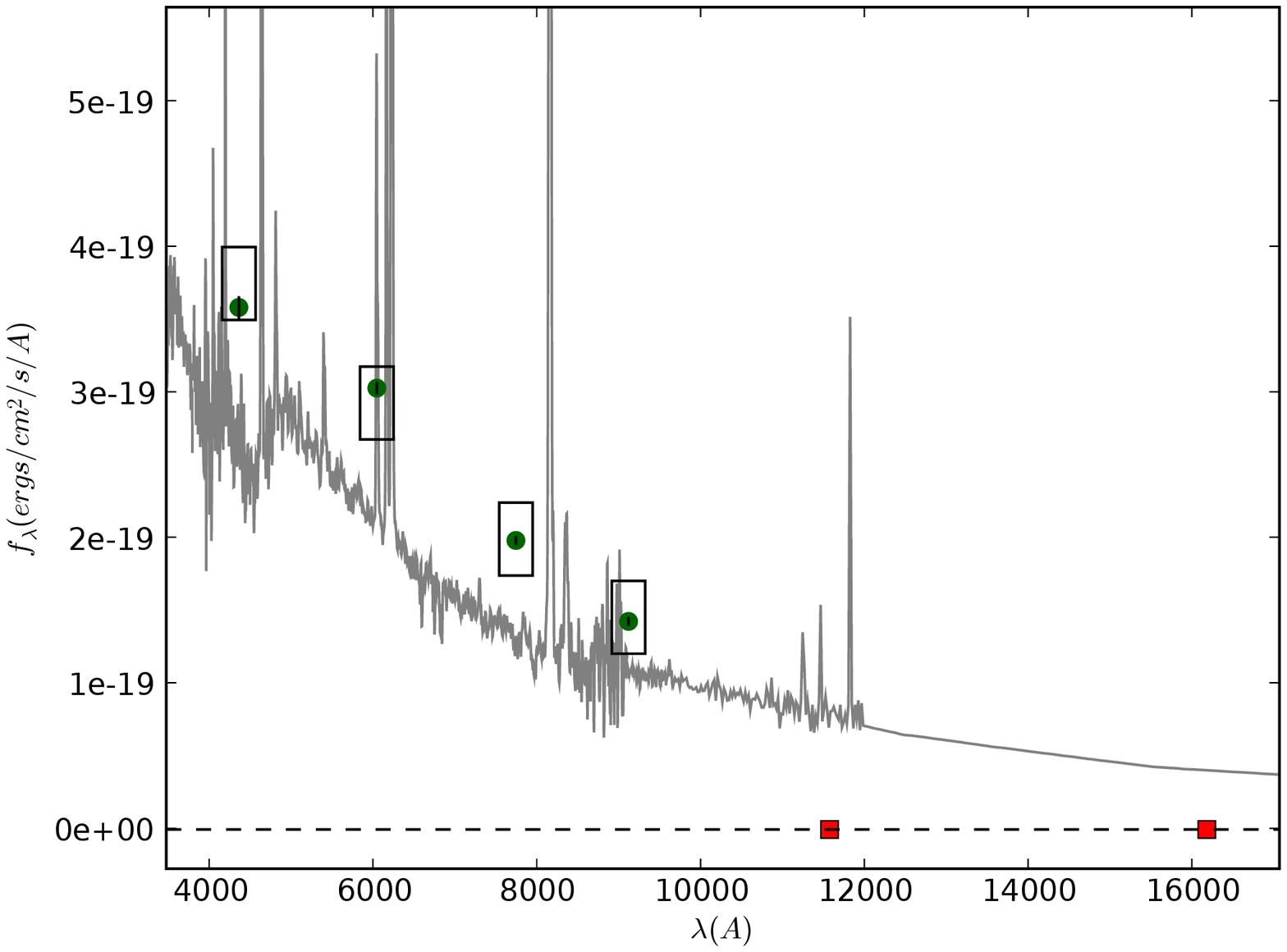}{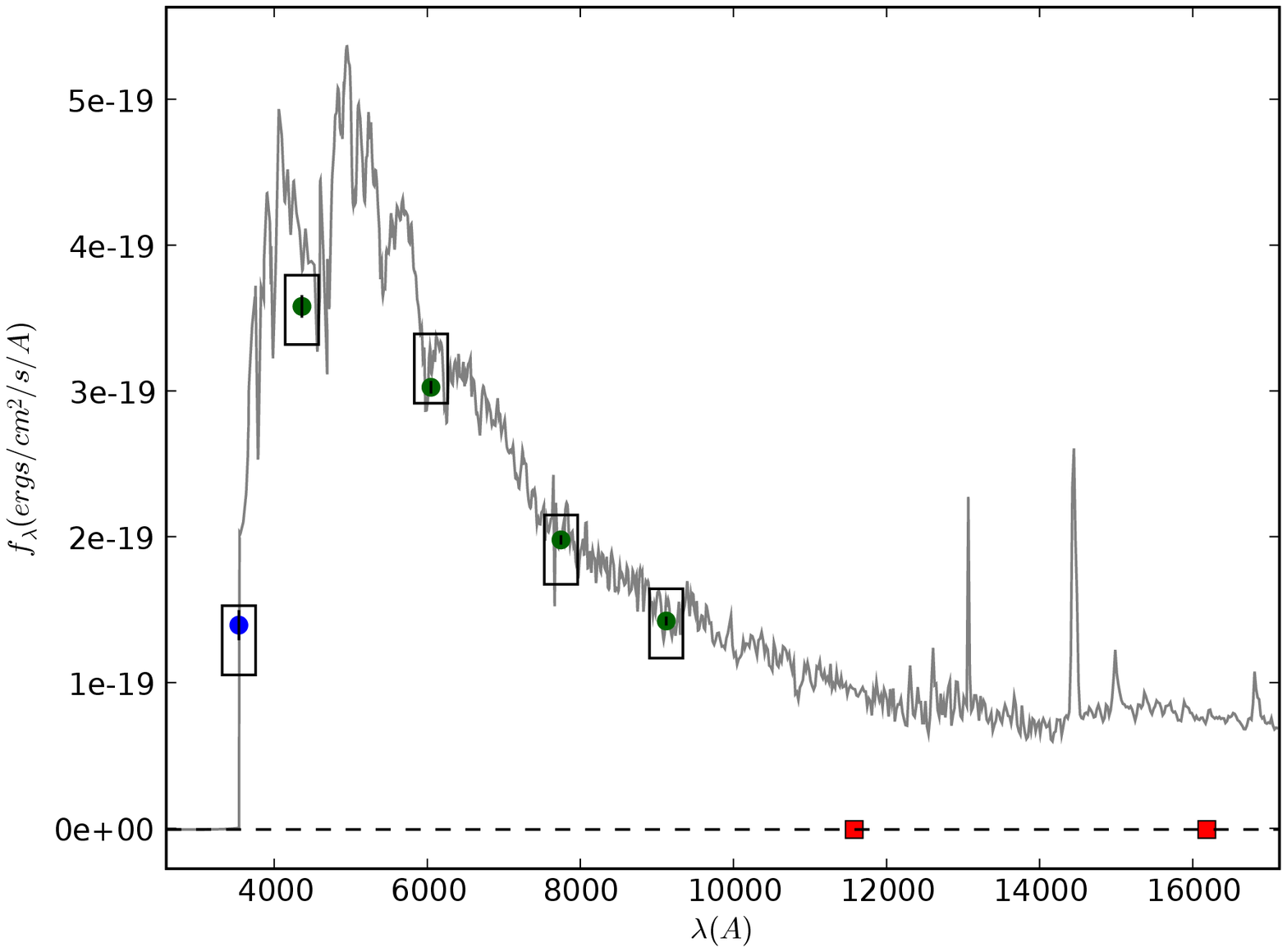}
\caption{Object 97: An example of a galaxy with $z_{\mathrm{spec}} = 2.69$ \citep{Popesso:2009p6629} 
that changes photo-$z$'s with the addition of the $u$-band. The left-hand side figures shows the 
photo-$z$ fit without $u$-band with z=0.24$_{-0.14}^{+2.67}$ and the right-hand side figures show the fit with $u$-band with z=2.90$\pm0.38$.
The addition of the $u$-band rules out the possibility of this being a low-redshift galaxy,
and therefore $P(z)$ only has one significant peak.
\label{obj97}}
\end{figure*}

\subsubsection{Improvement of photo-$z$'s with the Addition of the $u$-band}

A comparison of the redshift interval ($2.5\lesssim z \lesssim 3.5$) in the two panels of Figure \ref{photzspec}
shows a significant improvement in the photo-$z$'s with $u$-band data. The left panel depicts photo-$z$'s with 
$u$-band data while the right panel depicts photo-$z$'s without $u$-band data.
The $u$-band helps prevent catastrophic redshift errors that can occur because of a similarity
in the colors of low-redshift galaxies and high-redshift galaxies.
\citep{Ellis:1997p3771, FernandezSoto:1999p2784, Benitez:2000p3572}.
Usually, this degeneracy causes $P(z)$ to have multiple peaks representing
both possible redshifts in the absence of $u$-band data \citep{Coe:2006p1519}.
As discussed in \S4.1.2, however, sometimes secondary peaks are absent yielding
incorrect redshift uncertainties and possibly incorrect redshifts. 
For example, three galaxies (IDs 830, 4267, 5491) with $\chi^2_{\mathrm{mod}} < 4$ and $\tt ODDS$$> 0.99$ 
have catastrophic redshift errors without $u$-band photometry and gain accurate 
photo-$z$'s based on their spec-$z$'s after we include the $u$-band (e.g. see Figure \ref{obj830}). 
As expected, the u data clearly improve the photo-$z$'s of galaxies at $z\sim3$ when compared to the spec-$z$'s. 
However,  we are constrained to discussing small number statistics, and the true contamination fraction is unknown.

We would ideally like to compare a larger number of $z\sim3$ photo-$z$'s to spec-$z$'s to get a better understanding of the improvement
of the photo-$z$'s with the addition of the $u$-band, although such spectroscopic data are lacking in the UDF. 
However, since the $u$-band samples the Lyman break of $z\sim3$ galaxies, and we know that the photo-$z$'s with
$u$-band are more reliable than those without, we can investigate the changes in the photo-$z$'s. 
We therefore compare all the photo-$z$'s with $\chi^2_{\mathrm{mod}} < 4$ of galaxies with and 
without the $u$-band data in Figure \ref{photzcomp}, which highlights two effects. 
The first is that $P(z)$ changes markedly for 125 galaxies, mostly in the redshift interval 
$2\lesssim z \lesssim 3$, where the $u$-band probes the Lyman break. Of these, 102 change
their photo-$z$'s from $z\leq1$ to $z\sim2-3$, whereas 23 switch in the other direction. 
This change is a result of the code selecting different $P(z)$ peaks 
as the most probable redshift for galaxies having multiple peaks . 

The second effect shown in Figure \ref{photzcomp} is the removal of the degeneracy of 
$z\sim3$ and $z\sim0.2$ photo-$z$'s, due to the removal of one of the peaks in $P(z)$. 
One hundred seventy five galaxies went from being degenerate to non-degenerate, and are marked by the red crosses in Figure \ref{photzcomp}. 
Not all of the degenerate photo-$z$'s are removed; however, 
51 galaxies are degenerate between these redshifts as marked 
by the blue crosses in Figure \ref{photzcomp}, of which 17 were not degenerate before the addition of the $u$-band. 
The new degeneracies occur when the $u$-band best fit redshift is different compared to the best-fit to the other bands. The old degeneracies that are not removed
occur when the $u$-band does not conclusively rule out another template, often because they are faint.
Figure \ref{photzcomp} also includes galaxies with $\chi^2_{\mathrm{mod}} < 4$ in the photometric redshift fits selected to 
be at $z\sim3$ using the color selection method described in \S4.2 (below), which is useful when comparing the two methods. 
Figure \ref{obj97} shows an example of both effects, where the $u$-band changes the photo-$z$ and removes a secondary peak in $P(z)$ for a high-redshift galaxy. 

Out of 1384 galaxies with $\chi^2_{\mathrm{mod}} < 4$, there are 274 galaxies that have photo-$z$'s 
in the interval $2.5 \leq z \leq 3.5$ without $u$-band. The addition of the $u$-band 
increases this number by 91, to 365 galaxies that have a photo-$z$ in this redshift interval either 
with or without the $u$-band (including the 23 that switched to low redshift). 
Of these 365 galaxies, 161 galaxies either had their photo-$z$ changed or the 
degeneracy removed with the addition of the $u$-band. 
This shows that the addition of the $u$-band significantly changed the 
photo-$z$'s of the $z\sim3$ galaxy sample by $\sim 50\%$.

\subsection{Color Selection}

Color selection is an efficient means to select high-redshift galaxies, 
and extensive research has been carried out to determine the best color criteria to minimize
the interloper fraction from low-redshift galaxies or stars, e.g.
\citet{Steidel:1996p5981, Steidel:1996p5985,Steidel:1999p4108,Steidel:2003p1769,Adelberger:2004p3895, Cooke:2005p484}. 
The color selection criteria used in these studies are based on predicted colors of model star-forming galaxies at high redshift, 
which are then confirmed with spec-$z$'s, that result in known contamination fractions 
between 3\%--5\% \citep{Steidel:2003p1769, Reddy:2008p4837}.  Such low contamination fractions are achieved by avoiding colors
where low-redshift galaxies reside. 
While color selection techniques do not provide a complete sample of LBGs, they do an excellent job of selecting 
galaxies in a specific redshift range, as evidenced by their contamination fractions. 
While the UDF data provide an extraordinary data set, they also use a different set of filters than
used in previous color selection studies, meaning we must define new color criteria for LBG selection.
We therefore develop and calibrate new color criteria for selecting $z\sim3$ LBGs using the same methodology. 
Since our motivation is to generate a sample of galaxies to put constraints on the 
star formation efficiency at high redshift, we choose our color selection criteria
to best minimize possible low-redshift interlopers (see below).

\subsubsection{Color Selection Criteria}

Using the same approach as 
\citet{Steidel:1996p5981, Steidel:1996p5985,Steidel:1999p4108,Steidel:2003p1769}, \citet{Adelberger:2004p3895}, and \citet{Cooke:2005p484},
we derive galaxy colors by evolving different
galaxy SED templates to high redshift convolved with the total throughput of the different filters shown in Figure \ref{filter}.
We include galaxy SED templates consistent with our photometric redshifts described in \S4.1, 
with galaxy SED templates from \citet{Kinney:1996p6459},  \citet{Coleman:1980p4084}, and \citet{Bruzual:2003p4897}.
In addition we use a 2.0 Gyr Elliptical Galaxy from 
the \citet{Bruzual:2003p4897} synthesis code (E2G) since it is quite different than the elliptical 
galaxy SED template from \citet{Coleman:1980p4084} and represents possible low-redshift galaxies we wish to avoid. 
We apply the $K$-correction for different redshifts and correct for the opacity from the intergalactic medium by using
estimates from \cite{Madau:1995p4114}. 
The resultant colors and redshifts of the SED template galaxies are used to determine the appropriate color criteria to 
maximize the number of LBG candidates at $z\sim3$ while minimizing the contamination from objects at other redshifts. 
We test multiple color--color planes and find that for our set of filters,  $z\sim3$ LBGs can best be selected in a
($u-V$) versus ($V-z^\prime$) color--color plane.
Figure \ref{colorcut} plots the expected colors of different model galaxies at different redshifts in the ($u-V$) versus ($V-z^\prime$) diagram. 

The region defining candidate $z\sim3$ LBGs is indicated with the dashed black line in Figure \ref{colorcut}.
In order to avoid selecting low-redshift galaxies, this selection region 
excludes SED template colors for $z\le2.5$. The deviations from SED templates and 
photometric errors will cause intrinsic scatter in the color--color plane. We therefore leave $\sim0.2$ mag between
our color selection and the low-redshift elliptical galaxy SEDs that cause the largest contamination for galaxies at $z\sim3$.
In addition to the cut in the ($u-V$) versus ($V-z^\prime$) diagram, we also
apply secondary color cuts using the ($u-B$) color to improve our 
color selections by removing potential interlopers. 
We also include a cut on $V$-band magnitude, where the bright end does not remove any LBG candidates and the faint end
is the $V$-band magnitude determined in \S3.5 to keep the S/N of the $u$-band $>3\sigma$.
The following conservative constraints are used to select LBG candidates:
\begin{equation}
(u-V) \geq 1.0,
\label{eq:uv}
\end{equation}
\begin{equation}
(u-B) \geq 0.8,
\label{eq:ub}
\end{equation}
\begin{equation}
(V-z^\prime) \leq 0.6,
\label{eq:vz}
\end{equation}
\begin{equation}
3 (V-z^\prime) \leq (u-V) -1.2,
\label{eq:uvz}
\end{equation}
\begin{equation}
23.5 \leq V \leq  27.6.
\label{eq:vmag}
\end{equation}

\subsubsection{Reliability of Color Selection}

In order to test our selection criteria, we compare our selection of the 100 galaxies 
with reliable spec-$z$'s as described in \S4.1.3 and Table \ref{tab3}.
These spectra allow us to test the efficacy
of selecting targets via the ($u-V$) versus ($V-z^\prime$) color plane (Figure \ref{colorspec}).
No $z<2.5$ galaxies with reliable spec-$z$'s passed our color cut
confirming that our cut effectively excludes low-redshift galaxies.
We note that there are three $z>2.5$ objects that do not meet our color criteria in Figure \ref{colorspec}. 
The $z=3.68$ object (triangle, object 865) is 
classified as a quasar by \citet{Szokoly:2004p4004} due to active galactic nuclei (AGNs) activity. The $z=4.77$ object with ($V-z^\prime$)
is a bright $V$-band dropout galaxy, whose $V$-band magnitude (27.26) is 
just bright enough to remain in the sample
before the color cuts. Finally, the $z=3.80$ is also excluded by our color cut due to a decrease in the $V$-band magnitude, reddening the color.

There are a number of objects that could contaminate our sample of
 $z\sim3$ LBGs because their spectra are unusual and thus do not match
our model SEDs. While at brighter magnitudes the color selection criteria include stars and AGNs,
for our V $>$ 23.5 sample the color selection criteria are mainly 
contaminated by low-redshift galaxies at $z\lesssim0.2$ \citep{Reddy:2008p4837}.
Additionally, about $1/3$ of the Distant Red Galaxies (DRGs) fall
within the color selection sample of Steidel \citep{vanDokkum:2006p8396}, which is similar to our criteria. 
DRGs are galaxies at $z\gtrsim$2 that have faint UV luminosities, 
have previously undergone their episode(s) of star formation \citep{Franx:2003p8348},
and have stellar masses $\gtrsim10^{11}M_{\odot}$ \citep{vanDokkum:2004p8401, vanDokkum:2006p8396}.
However, given the DRGs' estimated space density of 
$(2.2\pm0.6)\times10^{-4}$ Mpc$^{-3}$ \citep{vanDokkum:2006p8396}, \citet{Reddy:2009p6997} conservatively determine that the 
fractional contribution would be $\sim$2\% for UV-faint sources such as our sample.

\begin{figure}
\plotone{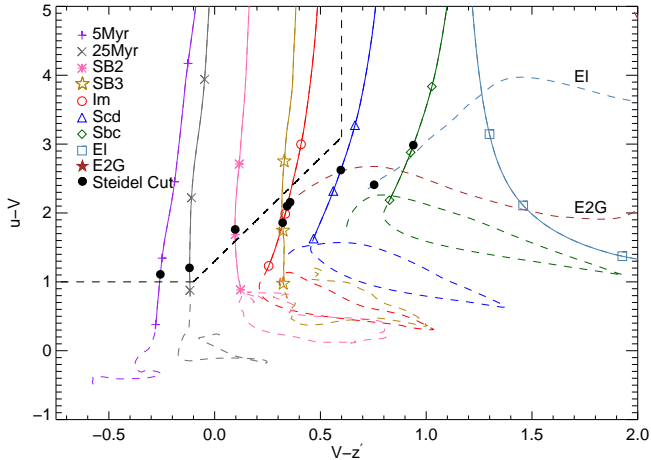}
\caption{($u-V$) versus ($V-z^\prime$) color--color diagram depicting different model galaxies at different redshifts, 
where the solid lines are $z>2.5$, the dashed lines are  $z<2.5$, and the symbols represent redshifts of 
2.5, 2.8, 3.0, and 3.2 which increase with increasing ($u-V$) color. 
We include galaxy SED templates from \citet{Kinney:1996p6459} for star bursting galaxies with different reddening (SB2, SB3),  
from \citet{Coleman:1980p4084} for 
star-forming galaxies (Im), spiral galaxies (Scd, Sbc), and elliptical galaxies (El). In addition, we use
and two faint blue galaxy SEDs with ages of 25 and 5 Myr and 
metallicities of Z=0.08 (25Myr, 5Myr), and a 2.0 Gyr Elliptical Galaxy (E2G) from 
the \citet{Bruzual:2003p4897} synthesis code. The area above and to the left of the dashed black line is the region selecting candidate LBGs. 
The filled black circles are the borderline colors for each SED template that correspond to the 
$z\sim3$ color cut by \citet{Steidel:2003p1769}, as described in \S4.2.2.
\label{colorcut}}
\end{figure}

\begin{figure}
\plotone{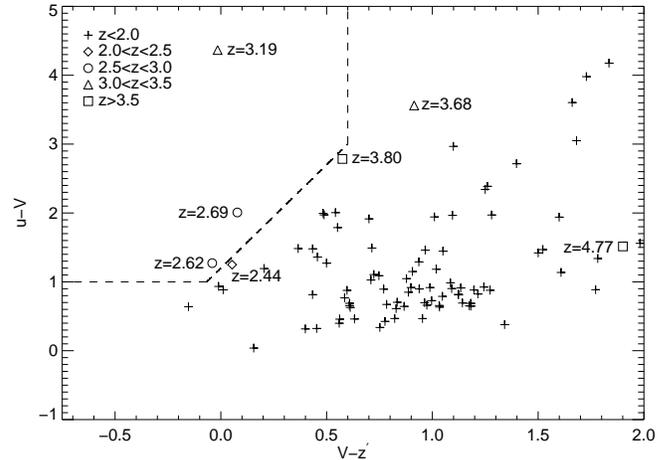}
\caption{Comparison of our selection criteria to 100 reliable spec-$z$ galaxies as described in \S4.1.3.
Object redshifts are binned and represented with different symbols, 
where crosses are for objects with $z<2.0$, 
diamonds for $2.0<z<2.5$, open circles for $2.5<z<3.0$,
triangles for $3.0<z<3.5$, and squares for $z>3.5$. The dashed line depicts our color cut. 
\label{colorspec}}
\end{figure}

As discussed earlier, to get a sense of the number of possible interlopers, we would like to have a large number of spectra to determine the contamination fraction. 
However, we only have a relatively small number of spectra (100) as shown in Figure \ref{colorspec} and Table \ref{tab3}. 
Instead, we compare our color cuts to the ($Un-G$) versus ($G-R$) color cuts
of \citet{Steidel:2003p1769} for redshifts $2.7\leq z \leq3.4$, and assume that our sample has similar contamination. 
In Figure \ref{colorcut}, we determine the redshift that 
corresponds to Steidel's color cut for a given SED template in ($Un-G$) and ($G-R$).
We then determine the ($u-V$) and ($V-z^\prime$) colors that this SED template has at this redshift
and compare those resulting colors to our color criteria by marking them with filled black circles in Figure \ref{colorcut}. 
The filled black circles are outside or near the edge of our 
color selection criteria, which shows that we are more conservative in our cut for the redder SED templates and about equivalently 
conservative for the bluer ones as the cut in \citet{Steidel:2003p1769}.
We therefore infer that our contamination fraction is comparable to those of 
\citet{Steidel:2003p1769} and \citet{Reddy:2008p4837}, where the contamination fraction of $z\sim3$ LBGs is $\sim3$\% for objects $23.5 \leq V \leq 25.5$ mag.

\section{Sample of z $\sim$ 3 Galaxies}

Photometric redshifts and color selection are both good ways to select $z\sim3$ galaxies. Photometric redshifts 
have the advantage of creating a larger sample since they can measure redshifts in regions of color space
that color selected samples avoid because of low-redshift galaxies. They also use more information than 
color selection, including all colors simultaneously to constrain the redshift. However, while the error rate of
the photo-$z$'s is not well defined, color selection is efficient, has a clearly defined contamination fraction, 
and allows direct comparisons to other studies. 
The completeness of our LBG selection is limited primarily on the $u$-band depth because the ACS 
bands from HST are deeper. In order to characterize this completeness, we compare our 
color selected LBGs with other studies in \S5.3.
To justify such comparisons, we compare the redshift distribution of this sample in \S5.1, and investigate our uncertainties from cosmic variance in \S5.2. 

In choosing our color selection criteria we chose to be conservative and create a less complete catalog 
with a small contamination fraction similar to that of \citet{Steidel:2003p1769} and 
\citet{Reddy:2008p4837} of $~$3\%. Depending on the purpose, a higher contamination fraction is acceptable 
in exchange for a larger and more complete sample. The photo-$z$ sample yields a more complete sample, and may even have a similar 
contamination fraction based on our spectroscopic sample, although due to our small numbers at the redshift of interest, it is not clear at this point. 
If the lowest contamination fraction possible is needed, 
a subset of $z\sim3$ LBGs that are both color selected and photo-$z$ selected are the most robust candidates available. 

\begin{figure}
\plotone{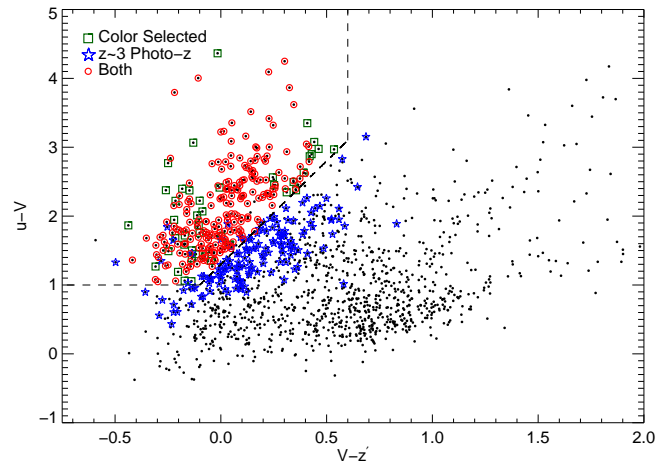}
\caption{All objects from table \ref{tab1} on a color--color diagram, where the dashed line 
refers to the selection criteria for constraints \ref{eq:uv}, \ref{eq:vz}, and \ref{eq:uvz} from \S4.2.1. 
There are 27 objects that are in the color selection area in this diagram
but are not color selected because of constraint \ref{eq:ub} from \S4.2.1. 
The 218 galaxies that are both color selected and photo-$z$ selected are marked by red circles, the 42 galaxies that are only color selected are marked by green squares, and 
the 147 galaxies selected only by photo-$z$'s are marked by blue stars, yielding a total of 407 $z\sim3$ LBG candidates. 
\label{objsel}}
\end{figure}

We present both samples of $z\sim3$ LBGs in Table \ref{tab4}, for a total of 407 candidates, along with their photo-$z$'s and colors. 
We distinguish the samples by designating them as either
color selected, photo-$z$ selected, or both. 
The photo-$z$ sample consists of galaxies in the redshift interval $2.5\leq z \leq 3.5$ that 
have $t_b>3$, $\tt ODDS$ $> 0.99$, and $\chi^2_{\mathrm{mod}} < 4$, and contains 365 galaxies. 
Of the 42 galaxies not included in our photo-$z$ sample, two 
have $z\sim2.2$, 11 have $z>3.5$, eight others have $\tt ODDS <0.99$, and 21 others have $\chi^2_{\mathrm{mod}} > 4$.
The color selected sample contains 260 galaxies, all of which have photo-$z$'s with $z>2$, with 258 that have $z>2.5$ and 11 that have $z>3.5$.
However, the overlap of the two samples is only 216 galaxies. 
We show the final LBG selection in Figure \ref{objsel}, which plots all objects from Table \ref{tab1} on a color--color diagram, 
with 287 galaxies falling in the color selection region corresponding to constraints \ref{eq:uv}, \ref{eq:vz}, and \ref{eq:uvz} from \S4.2.1.
There are 27 galaxies that are in the selection area in this diagram, but are rejected by the ($u-B$) color (constraint \ref{eq:ub}), 
leaving 260 galaxies that are color selected. The objects marked as blue stars that are selected by photo-$z$'s but not by color selection
are generally galaxies at $z\sim2.8$ that are missed by our color selection criteria in order to avoid elliptical and low-redshift galaxies.

\subsection{Redshift Distribution}

\begin{figure*}
\plottwo{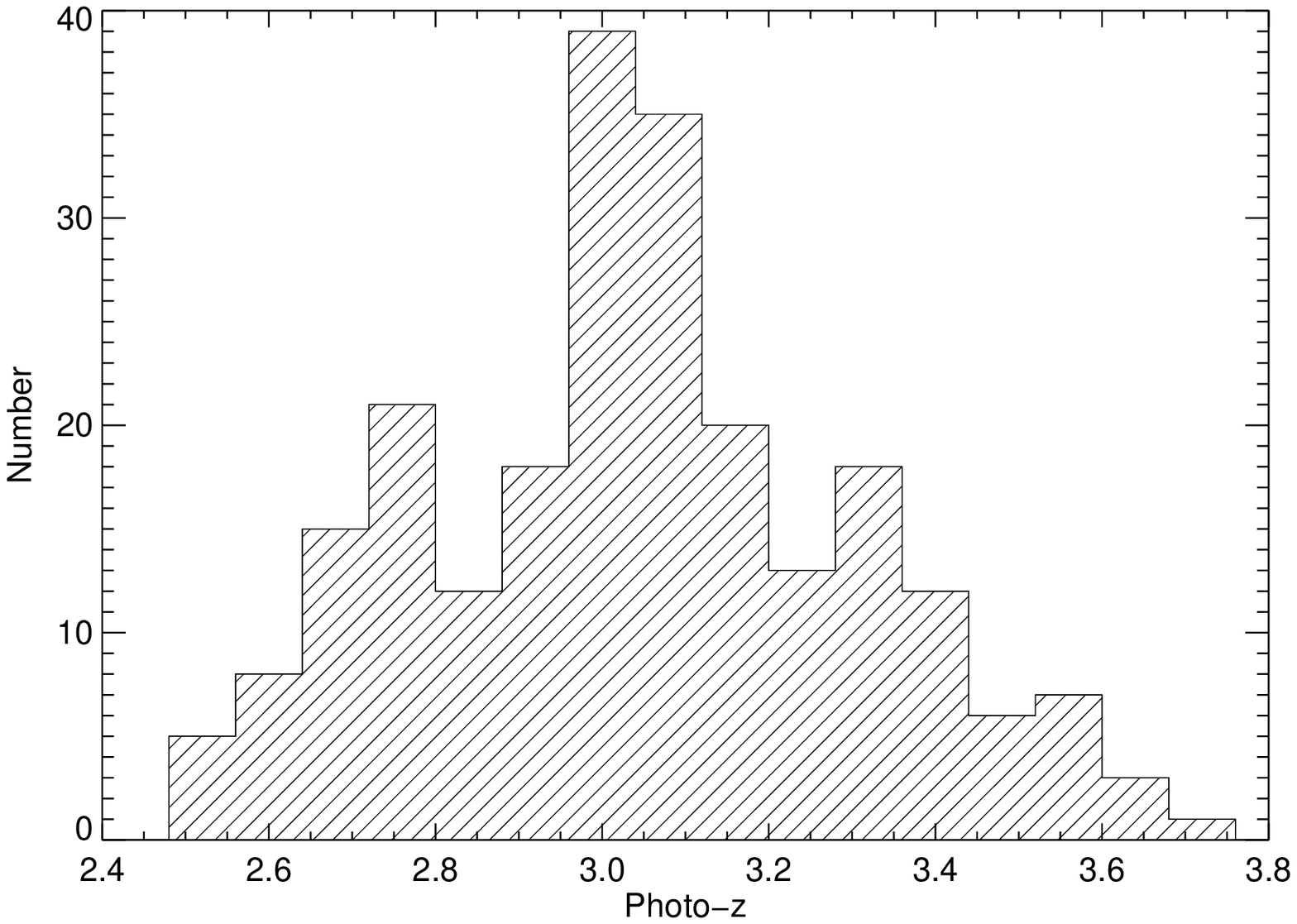}{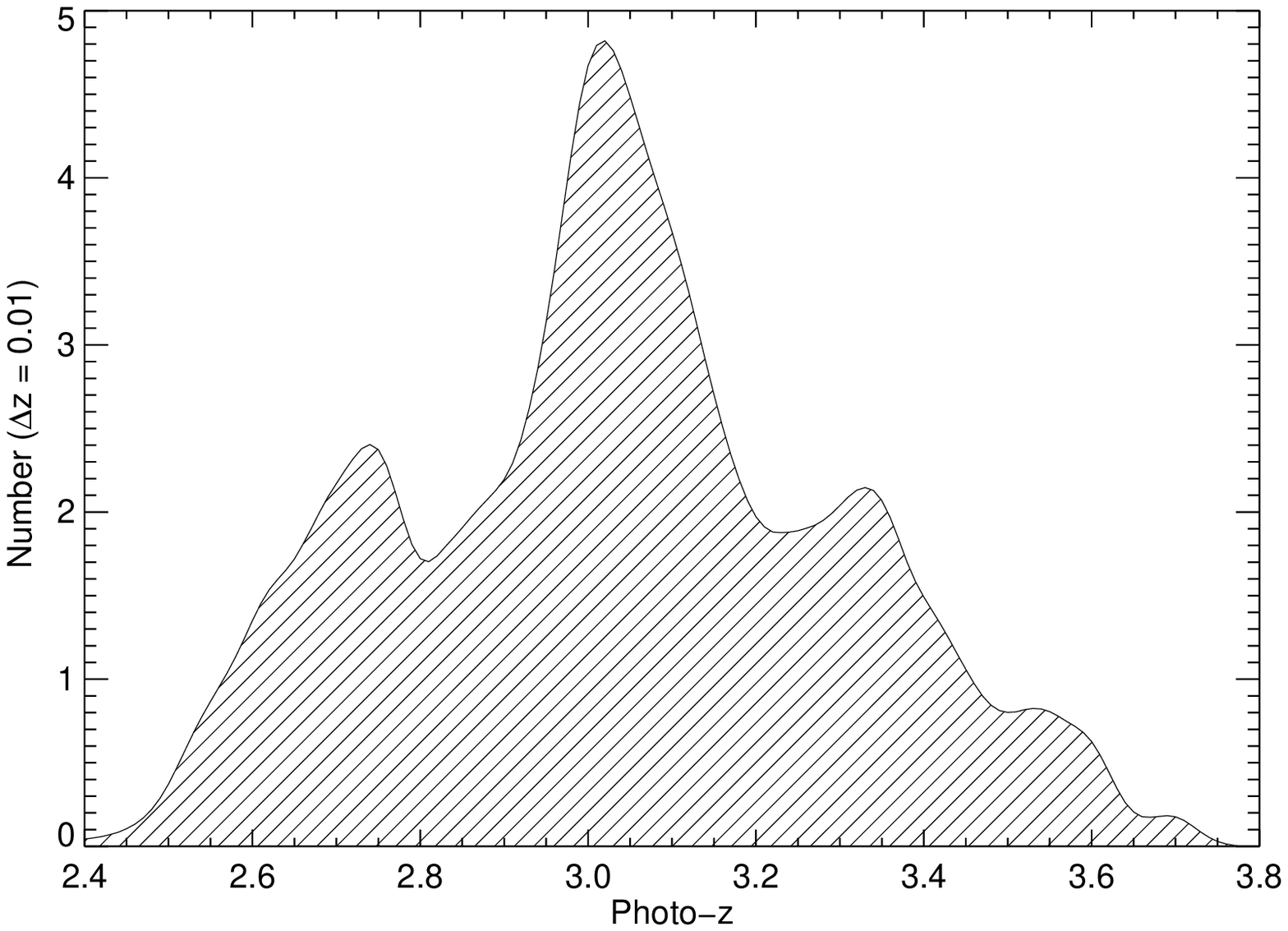}
\caption{Histogram of 235 color selected photometric redshifts with $\chi^2_{\mathrm{mod}} < 4$. 
The mean redshift is $3.0\pm0.3$, and the median uncertainty of the photo-$z$'s is $\pm$0.4.
The left panel is a traditional histogram, and the right panel is the sum of the probability histograms $P(z)$ of the individual galaxies.
The redshift distribution selected are similar to those of \citet{Steidel:2003p1769}.
\vspace{0.2in}
\label{zhist}}
\end{figure*}

In order to understand the redshift distribution of the color selected LBG sample, we look at the photo-$z$'s that meet the color
selection criteria and our photo-$z$ criteria of $\chi^2_{\mathrm{mod}} < 4$. This leaves a sample of 235 galaxies that have a 
mean redshift of $3.0\pm0.3$ in the redshift interval $2.4 \lesssim z \lesssim 3.8$ (see left panel in Figure \ref{zhist}),
with a median uncertainty in the photo-$z$'s of $\pm$0.4. 
Additionally, we investigate the redshift distribution by adding the probability histograms $P(z)$ of the individual galaxies, which
yields a similar result (see the right panel in Figure \ref{zhist}).
The redshifts selected are similar to those of \citet{Steidel:2003p1769},
with $\sim$80\% of our sample matching their reported redshift interval $2.7<z<3.4$. In fact, their distribution is very similar to ours,
with a number of their LBGs falling outside of this interval. Given the large uncertainties in our photo-$z$'s, 
we conclude that our color selected redshift distribution is similar to that of \citet{Steidel:2003p1769} within our uncertainties.
The similar redshift distribution of our color selected sample justifies our comparison of the number densities of LBGs in \S5.3.

\begin{deluxetable*}{crccccccc}
\tabletypesize{\scriptsize}
\tablecaption{Catalog of Candidate z$\sim$3 LBGs in the UDF
\label{tab4}}
\tablewidth{0pt}
\tablehead{
\colhead{ID\tablenotemark{a}} &
\colhead{$z_b$\tablenotemark{b}} &
\colhead{$\chi^2_{\mathrm{mod}}$\tablenotemark{c}} & 
\colhead{$\tt ODDS$\tablenotemark{d}} &
\colhead{Type\tablenotemark{e}} &
\colhead{$V$ (mag)} &
\colhead{$u-V$ (mag)} &
\colhead{$u-B$ (mag)} &
\colhead{$V-z^\prime$ (mag)}}

\startdata

\vspace{0.03in}
22 &3.23$_{-0.41}^{+0.41}$ &0.44 &1.00 &2 &27.38$ \pm $0.03 &1.53$ \pm $0.36 &0.96$ \pm $0.36 &0.49$ \pm $0.04 \\
\vspace{0.03in}
76 &2.62$_{-0.35}^{+0.36}$ &0.52 &1.00 &2 &26.47$ \pm $0.02 &1.39$ \pm $0.16 &1.19$ \pm $0.16 &0.11$ \pm $0.03 \\
\vspace{0.03in}
84 &3.11$_{-0.40}^{+0.40}$ &0.01 &1.00 &1 &26.56$ \pm $0.02 &2.40$ \pm $0.35 &1.69$ \pm $0.35 &0.03$ \pm $0.04 \\
\vspace{0.03in}
97 &2.87$_{-0.38}^{+0.38}$ &0.13 &1.00 &1 &25.02$ \pm $0.01 &2.01$ \pm $0.06 &1.48$ \pm $0.07 &0.08$ \pm $0.01 \\
\vspace{0.03in}
99 &2.81$_{-0.37}^{+0.37}$ &0.32 &1.00 &2 &24.74$ \pm $0.01 &1.96$ \pm $0.07 &1.51$ \pm $0.08 &0.52$ \pm $0.01 \\
\vspace{0.03in}
101 &2.66$_{-0.36}^{+0.36}$ &0.17 &1.00 &2 &27.38$ \pm $0.02 &1.45$ \pm $0.32 &1.09$ \pm $0.32 &0.09$ \pm $0.05 \\
\vspace{0.03in}
131 &3.06$_{-0.40}^{+0.40}$ &0.21 &1.00 &1 &26.95$ \pm $0.02 &2.02$ \pm $0.35 &1.43$ \pm $0.36 &0.03$ \pm $0.05 \\
\vspace{0.03in}
209 &2.67$_{-0.36}^{+0.36}$ &9.03 &1.00 &3 &26.92$ \pm $0.02 &1.27$ \pm $0.34 &1.26$ \pm $0.34 &-0.31$ \pm $0.05 \\
\vspace{0.03in}
213 &3.14$_{-0.41}^{+0.41}$ &0.18 &1.00 &2 &26.91$ \pm $0.02 &1.95$ \pm $0.34 &1.19$ \pm $0.35 &0.56$ \pm $0.04 \\
\vspace{0.03in}
230 &3.70$_{-0.46}^{+0.46}$ &0.00 &1.00 &3 &25.68$ \pm $0.01 &3.08$ \pm $0.35 &1.29$ \pm $0.36 &0.44$ \pm $0.02

\enddata

\vspace{0.03in}

\tablecomments{Table \ref{tab4} is published in its entirety in a machine-readable form in the online version of the Astrophysical Journal. 
A portion is shown here for guidance regarding its form and content. 
Photometric redshifts should only be used if they have low $\chi^2_{\mathrm{mod}}$ and good $\tt ODDS$ values as described in \S4.1.1.
$V$ magnitudes are total AB magnitudes, and colors are isophotal colors. All photometry other than the u-band are from \citet{Coe:2006p1519}.
Nondetections in u-band are given 3$\sigma$ limiting magnitudes. }
\tablenotetext{a}{ID numbers from \citet{Coe:2006p1519}.}
\tablenotetext{b}{Bayesian photometric redshift (BPZ)  and uncertainty from 95\% confidence interval.}
\tablenotetext{c}{Modified reduced chi-square fit, where the templates are given uncertainties.}
\tablenotetext{d}{Integrated $P(z)$ contained within $0.1(1+z_b)$.}
\tablenotetext{e}{Type of LBG selection, where 1=both color and photo-$z$ selected, 2=only photo-$z$ selected, and 3=only color selected.}

\end{deluxetable*}

\subsection{Cosmic Variance}

The UDF has a very small volume, with our overlap area consisting of 11.56 arcmin$^{2}$, 
which in the redshift interval $2.5\leq z \leq 3.5$ is a comoving volume of $\sim38000$ Mpc$^3$.
A single pointing with a small solid angle and such a small volume is likely to be affected by cosmic variance, 
yielding larger than Poisson uncertainties in the LBG number counts. 
It is important to estimate the cosmic variance effect in order to understand the systematic 
uncertainties for comparisons with number densities of LBGs in the literature.

We calculate the cosmic variance using the code from \citet{Newman:2002p7424} 
and the prescription from \citet{Adelberger:2005p4252}
to get the fractional error per count $\sigma/N$
for our given volume in the redshift interval $2.5\lesssim z \lesssim 3.5$ . 
The variance is determined from the integral of the 
linear regime of the cold dark matter (CDM) power spectrum ($P(k)$), 
\begin{equation}
\sigma^2_{\mathrm{CDM}} = \frac{1}{8\pi^3}\int P(k)|\tilde{W}(k)|^2d^3k  \; \; \; ,
\label{sigma}
\end{equation}
where $\tilde{W}(k)$ is the Fourier transform of our survey volume. Since we want the variance of galaxy counts rather than CDM fluctuations, 
we need to correct for the clustering bias $(b)$ of LBGs to get their variance ($\sigma_g^2$), 
where $\sigma_g^2 \simeq b^2\sigma^2_{\mathrm{CDM}}$. The galaxy bias for typical LBGs is then calculated from the 
ratio of galaxy to CDM fluctuations in spheres of comoving radius 8 $h^{-1} \mathrm{Mpc}$, 
where $\sigma_8(z)$ represents the CDM fluctuations for our redshift, and $b=\sigma_{8,g} / \sigma_8(z)$  \citep{Adelberger:2005p4252}. 
The resulting variance depends on a 
fit to the LBG correlation function  $\xi_g(r)=(r/r_o)^{-\gamma}$, where $r_o$ is
the spatial correlation length  and $\gamma$ the correlation index. 
The galaxy variance is then
\begin{equation}
\sigma^2_{8,g}=\frac{72(r_o/ 8\; h^{-1} Mpc)^\gamma}{(3-\gamma)(4-\gamma)(6-\gamma)2^\gamma}  \; \; \; ,
\label{sigma2}
\end{equation}
\citep[][eq. 59.3]{JamesEdwinPeebles:1980p8194} from which we can then calculate the fractional error per LBG count. 

Empirical fits to the correlation function yield differing values for $r_o$ and $\gamma$ depending on the sample, 
redshift distribution, luminosity range, and redshift,
which in turn affect the value of $\sigma^2_{8,g}$ and our fractional uncertainty
\citep{Adelberger:2005p4252, Hildebrandt:2007p4025, Hildebrandt:2009p8300, Kashikawa:2006p4666, 
Lee:2006p4240, Ouchi:2004p4668, Ouchi:2005p4177, Yoshida:2008p1419}. 
We avoid looking at samples covering a small area of the sky such as the HDF \citep{Giavalisco:2001p7613}, 
as such studies are also plagued by cosmic variance
as shown in \citet{Ouchi:2005p4177}. The values in the larger studies generally vary between $r_o \sim2.8-5.5$ and $\gamma \sim1.5-2.2$, with an
increasing  $r_o$ and $\gamma$ with increasing luminosity, i.e., the brighter LBGs are more strongly clustered. 
There is also some minor evolution with
redshift, where the higher redshift galaxies are more clustered \citep{Hildebrandt:2009p8300}. 
These values result in $\sigma^2_{8,\mathrm{LBGs}}$ of $\sim0.56--1.1$ which correspond to a fractional error per count of $\sim0.14-0.28$. Our
sample includes the fainter less clustered LBGs at $z\sim3$, so we are on the less clustered side of this range.
We therefore adopt a fractional error per count ($\sigma/N$) of $\sim$0.2 in the rest of this study, which suggests that we could detect a relatively
large over-density or under-density in our small volume. In fact, there is evidence of an over-density of $z\sim3$.7 galaxies in the Chandra Deep Field South, 
of which the UDF is a part \citep{Kang:2009p8043}. 
However, no clear over-density is indicated in the correlation length measured from the GOODs 
survey covering the same area on the sky \citep{Lee:2006p4240} at slightly higher redshift.
We use the above estimate of the cosmic variance for our small field of view to constrain our results of the number densities of LBGs in \S5.3.

\subsection{LBG Number Counts}

The number counts of $z\sim3$ LBGs per unit of magnitude indicate the completeness of the LBG selection, and can be 
compared to number counts from other studies. Such data are only available for color-selected samples, and we therefore
only use our color-selected sample in our comparison. 
The same comparison could be accomplished with the luminosity function, 
which we do not calculate because we only have one pointing and
our comoving volume is small. In other words, we have a small number of LBGs that
in conjunction with the uncertainty due to cosmic variance discussed above,
would not yield meaningful constraints on the luminosity function. 
Additionally, to calculate the luminosity function, Monte Carlo simulations need to be
run as described in \citet{Reddy:2008p4837} that are computationally prohibitive given our complex analysis technique to get
reliable photometry with largely varying PSFs \citep{Laidler:2007p2733}. We therefore compare our results
with the number counts from other studies. We stress that we are comparing number counts that are not corrected for completeness,
and therefore the counts will fall at faint magnitudes in each study due to sample incompleteness.

Our ground-based LRIS images have much greater PSF FWHMs ($\sim$1$\tt''$.3)  than
the HST $V$-band image ($\sim$0.09$\tt''$), thus affecting the observed number counts.
The dominant effect is caused by the blending of neighboring objects that affects the color of objects and therefore their selection. 
In addition, at significantly lower resolution, isolated compact faint objects have part of their flux lost to the noise floor of the background. 
This causes the faintest objects to go undetected in the low-resolution images, 
even if they would have been detected in a similarly sensitive high-resolution image.
We use the HST $V$-band image to determine our high-resolution-detection (HRD) 
number counts and correct for this resolution effect. We convolve the HST $V$-band image 
with the PSF modeled from the LRIS $V$-band image (FWHM of $\sim$1$\tt''$)
that was taken concurrently with the $u$-band data.
This yields a low-resolution-detection (LRD) image from which a segmentation map is generated using $\tt SExtractor$. 

\begin{figure}
\plotone{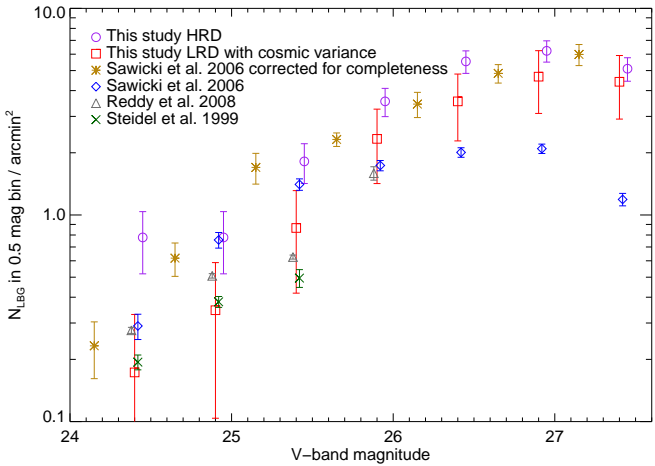}
\caption{Number counts of $z\sim3$ LBGs in 0.5 mag bins per square arcminute not corrected for completeness.
Magnitudes are slightly offset to make the figure more clear. The full number counts based 
on the high-resolution-detections are marked HRD and are purple circles. The LRDs
are marked LRD and are in red squares. The LRD are the number counts that would 
have been measured if the object detections would have been made using ground-based images with FWHM of 1$\tt''$.2. 
For the HRD sample, we include only Poisson uncertainties, while for the LRD we 
include both Poisson uncertainties and the cosmic variance estimate (see \S5.2).
Our number counts are plotted along with the studies of \citet{Reddy:2008p4837}, \citet{Sawicki:2006p1733}, 
and \citet{Steidel:1999p4108}. The only points corrected for completeness are the gold crosses for the \citet{Sawicki:2006p1733}
data. 
\label{numcounts}}
\end{figure}

The final photometry is measured with this new segmentation map in all bands using $\tt sexseg$ as discussed in \S3.4, 
and then the same color selection is used as discussed in \S4.2.1. 
Figure \ref{numcounts} shows the number counts per half magnitude bin per square arcminute for both the HRD and LRD, along
with number counts from \citet{Reddy:2008p4837}, the Keck Deep Field (KDF) \citep{Sawicki:2006p1733}, and \citet{Steidel:1999p4108}. 
For the HRD sample, we include only Poisson uncertainties for reference, while for the LRD we include both Poisson uncertainties and the
fractional error per count of 0.2 to take into account the cosmic variance (see \S5.2).
The figure shows that we are photometrically complete to $V\sim27$ mag, 
after which we start losing LBGs due to the sensitivity of the $u$-band image.  
In order to compare to other studies, we convert their $R$-band magnitudes from Keck to $V$-band magnitudes from HST with a $K$-correction 
at $z\sim3$ for the Im galaxy template described in \S4.1, which results in an R-V color of $\sim-0.15$.
The LBG number counts in Figure \ref{numcounts} of the LRD's agree at the brighter end to within our uncertainties, 
however, at the fainter end ($V>26.0$), our results are larger than the uncorrected KDF results, which is the only survey to probe 
to equivalent depths.

Based on the turnover in the LBG number counts in
Figure~\ref{numcounts}, the KDF and our LRD study appear to be
complete to $V\gtrsim26$ and $V\sim27$, respectively.  The cumulative
number counts for $V<26$ is $3.7\pm0.6$ LBGs
arcmin$^{-2}$  in our LRD study and $4.3\pm0.2$ LBGs arcmin$^{-2}$ for the KDF, where
the uncertainties are Poisson.  These results are consistent with each
other, without invoking cosmic variance, and suggest that the
difference in number counts for $V\gtrsim26$ is due to differing
completeness limits.

In order to understand this better, we also include the KDF number
counts corrected for LBG completeness in Figure \ref{numcounts}. 
This completeness correction is different than the one applied in
\citep{Sawicki:2006p1733}, as it previously included the volume correction
simultaneously.  The correction applied here assumes that all $z\sim3$
LBGs are well-represented by the colors of a fiducial LBG with (1)
fixed age of 100 Myr (2) a fixed redshift $z=3$, and (3)
\citet{Calzetti:1994p4914} dust extinction for $E(B-V)=0.2$. 
The incompleteness is then
calculated by planting objects with these colors in the KDF images and
determining how many are recovered, with the uncertainties estimated
via bootstrap resampling (Marcin Sawicki, private communication). 
This is not as careful a correction as applied in
\citep{Sawicki:2006p1733}, but serves to investigate the completeness
differences in our studies with respect to LBG detection.  Our LRD
study number counts are consistent with the KDF LBG completeness
corrected number counts down to $V\sim27$.  This helps reinforce that
we are likely complete in detecting $z\sim3$ LBGs to $V\sim27$
magnitude, making our study the deepest to date.

\section{Summary}

We use newly acquired ground-based $u$-band imaging with a depth of  30.7 mag arcsec$^{-2}$ ($1\sigma_{u}$ sky fluctuations) and
an isophotal limiting $u$-band magnitude of 27.6 mag to create a reliable sample of 407 $z\sim3$ LBGs in the UDF. 
We use the template-fitting method $\tt TFIT$ \citep{Laidler:2007p2733}
to measure accurate photometry without the need for aperture corrections, and obtain robust colors across the
largely varying PSFs of the UDF ACS images (0.$\tt''$09 FWHM) and the $u$-band image (1.$\tt''$3 FWHM). The results are as follows:

1. We calculate photometric redshifts for 1457 galaxies using the Bayesian algorithm of \citet{Benitez:2000p3572}, \citet{Benitez:2004p3578}, and \citet{Coe:2006p1519}, 
of which 1384 are reliable with $\chi^2_{\mathrm{mod}} < 4$. 
We find that the previous photo-$z$'s by \citet{Coe:2006p1519} 
do a good job of determining redshifts even without the $u$-band if their uncertainties are taken into account. However, these uncertainties are often 
quite large at $z\sim3$ due to the color degeneracy of $z\sim3$ and $z\sim0.2$ galaxies.

2. The $u$-band significantly improves $z\sim3$ photo-$z$ determinations: out of  1384 galaxies, 175 galaxies
no longer have degenerate photo-$z$'s, and 125 of the galaxies changed their primary photo-$z$ with the addition of the $u$-band.
In fact, the addition of the $u$-band changed the photo-$z$'s of $\sim50\%$ of the $z\sim3$ galaxy sample. 

3. We find that even when using the $u$-band photometry and restricting the sample of photo-$z$'s to those with good $\chi^2_{\mathrm{mod}}$,
catastrophic photo-$z$ errors can still occur, although they are rare (only 1 out of 93 
galaxies with spectroscopic redshifts and photometric redshifts with $\chi^2_{\mathrm{mod}} < 4$ and $\tt ODDS$ $ > 0.99$). 
We found \textit{no} catastrophic photo-$z$'s in the redshift interval of interest, ($2.5\lesssim z \lesssim 3.5$), although 
only five objects have spec-$z$'s in this interval. In contrast, three galaxies at $2.5\lesssim z \lesssim 3.5$ 
had catastrophic photo-$z$'s before the addition of the $u$-band.
The contamination fraction of the $z\sim3$ photo-$z$ sample is likely small as we sample the Lyman break for these galaxies, and show excellent
spectroscopic agreement. However, given the small numbers, the overall error rate of the photo-$z$'s is not well defined. 

\vspace{5.8in}

4. We find excellent agreement of our color selected sample with the spectroscopic $z\sim3$ sample, with no low-redshift galaxies
falling in our color selection area, confirming our chosen criteria.
We specifically choose a conservative color criteria that are similar to the cuts of \citet{Steidel:2003p1769} such that
we can infer that the $\sim3\%$ contamination fraction for $z\sim3$ LBGs of \citet{Steidel:2003p1769} and \citet{Reddy:2008p4837} applies to our data set.

5. The completeness of our LBG selection depends largely on the $u$-band depth because the ACS 
bands from HST are deeper. In order to characterize this completeness, we compare our 
color selected LBGs with other studies and find that we present the deepest sample of $z\sim3$ LBGs currently available, likely complete to $V\sim27$ mag. 

This reliable sample of $z\sim3$ LBGs can be used to further the studies of LBGs and star formation efficiency of gas at $z\sim3$ 
through the most sensitive high-resolution images ever taken; the Hubble Ultra Deep Field. 

\vspace{0.7in}


The authors thank, Dan Coe, Victoria Laidler, Eric Gawiser, and Alison Coil for valuable discussions, 
Marcin Sawicki for providing the KDF number counts,
and Narciso Benitez for providing re-calibrated SED templates for BPZ.
Support for this work was provided by NSF grant AST 07-09235.
J. C. acknowledges generous support by Gary McCue.
The W. M. Keck Observatory is operated as a scientific partnership among the California Institute 
of Technology, the University of California and the National Aeronautics and 
Space Administration.  The Observatory was made possible by the generous 
financial support of the W. M. Keck Foundation.  The authors recognize and 
acknowledge the very significant cultural role and reverence that the summit 
of Mauna Kea has always had within the indigenous Hawaiian community.  We are 
most fortunate to have the opportunity to conduct observations from this 
mountain.  

{\it Facility:} 
 \facility{Keck:I (LRIS)}, \facility{HST (ACS, NICMOS)}

\bibliography{lbgbib}

\end{document}